\documentclass[12pt]{iopart}

\usepackage[colorlinks=true,linkcolor=blue,urlcolor=blue,citecolor=blue]{hyperref}
\usepackage{xspace}
\usepackage{url}
\usepackage{graphicx}
\usepackage{amsmath,amssymb,bm}

\begin{document}

\title{Enhanced directionality of active processes in a viscoelastic bath}

\author{Biswajit Das$^1$, Shuvojit Paul$^1$, Sreekanth K. Manikandan$^2$, and Ayan Banerjee$^1$}
\address{$^1$Department of Physical Sciences, Indian Institute of Science Education and Research Kolkata, Mohanpur Campus, Mohanpur, West Bengal 741246, India.  $^2$Department of Chemistry, Stanford University, Stanford, CA, USA 94305. }
\ead{\mailto{bd18ip005@iiserkol.ac.in}, \mailto{shuvojit.pdf@iiserkol.ac.in}, \mailto{sreekm@stanford.edu}, and \mailto{ayan@iiserkol.ac.in}.}

\vspace{10pt}
%\begin{indented}
%\item[]June 2023
%\end{indented}

\begin{abstract}
Active fluctuations are known to play a significant role in the intracellular transport of passive objects. However, the effect of viscoelasticity of the environment in shaping such processes is relatively less understood. Here, with a minimal experiment using a driven colloid in a viscoelastic bath, we show that viscoelasticity significantly increases the mean injected power to the passive object ($\sim 50\%$ compared to a viscous medium), for the same strength of the external driving. Additionally, we observe a notable reduction in negative work fluctuations across a wide range of driving amplitudes. These findings collectively suggest an enhanced directionality in driven processes within a viscoelastic bath, which we attribute to the emergence of interactions between the colloid and the viscoelastic medium.
\end{abstract}

%
% Uncomment for keywords
%\vspace{2pc}
%\noindent{\it Keywords}: XXXXXX, YYYYYYYY, ZZZZZZZZZ
%
% Uncomment for Submitted to journal title message
%\submitto{\NJP}
%
% Uncomment if a separate title page is required
%\maketitle
% 
% For two-column output uncomment the next line and choose [10pt] rather than [12pt] in the \documentclass declaration
%\ioptwocol
%
\section{Introduction}

 Microscopic biophysical processes are vital for the functioning of living organisms. Many of these processes, such as intracellular transport of passive objects \cite{almonacid2015active,mogre2020getting}, are non-equilibrium in nature, mostly driven by ATP-consuming molecular motors such as kinesins \cite{hirokawa2009kinesin,balint2013correlative}. Developing accessible models for these processes is of utmost importance as it provides novel perspectives into the functioning of the fundamental components of life within complex environments.

In a zeroth order approximation, such processes can be modelled assuming a separation of time-scales between the environmental and system degrees of freedom, leading to, effectively, a Markovian description for the dynamics of the system. The thermodynamics of such processes has been extensively studied in recent decades, particularly in the context of stochastic thermodynamics \cite{seifert2008stochastic, seifert2012stochastic}. This field has focused on various aspects, including fluctuation relations \cite{seifert2005entropy, jarzynskia2008nonequilibrium, jarzynski2011equalities}, as well as advancements in stochastic inference techniques \cite{seifert2019stochastic, manikandan2020inferring, van2020entropy, otsubo2020estimating, manikandan2021quantitative, otsubo2022estimating, das2022inferring, barato2015thermodynamic, horowitz2020thermodynamic}. These theories have now been well tested in experiments, and have provided us with accurate estimates of thermodynamic quantities, such as equilibrium free energy differences and dissipation for specific biophysical systems \cite{liphardt2002equilibrium,toyabe2010nonequilibrium} and efficiencies of molecular motors \cite{kinosita2000rotary,ariga2018nonequilibrium}. 

A vast majority of these experiments are carried out in \textit{in-vitro} conditions where the experimental medium is usually a homogeneous viscous solution and the Markovian assumption is valid. This is also the case for most artificially synthesized molecular motors, Brownian Carnot engines and other nanoscale devices \cite{ariga2018nonequilibrium,martinez2016brownian}. Contrary to this, in reality, microscopic biophysical processes differ from simplified models by two key features. Firstly, living cells possess inherent activity and dynamic behavior, contributing to complex fluctuations that primarily govern the underlying processes, except in some situations where passive diffusion also may play big roles. For example, diffusive processes with directed motility of various intracellular elements are crucial in positioning and distributing organelles inside a cell \cite{lin2016active,witzel2019heterogeneities}. However, overall, the hidden nature of many such sub-cellular processes leads to non-Markovian dynamics for the observable microscopic processes. Secondly, the intracellular medium is highly viscoelastic. This viscoelasticity arises from the intricate composition and structure of the cellular environment, characterized by a combination of viscous and elastic properties \cite{guigas2007probing,bhat2012viscoelasticity}. This also leads to memory effects due to the large relaxation times, which again breaks the effective Markovian description for the dynamics of the system \cite{wilson2011microrheology,demery2014generalized,gomez2015transient,doi1988theory,paul2019single,kundu2021single}. 

Quite naturally, both of these issues have individually generated significant recent interest. Studies on various systems, such as bacterial suspensions or self-propelled particles, have revealed that active fluctuations can induce non-trivial transport phenomena inside cells \cite{colin2020active,xie2022contribution}, enhance mixing \cite{sokolov2009enhanced} and diffusion \cite{brangwynne2009intracellular,caspi2000enhanced}, and even drive the emergence of collective behaviour \cite{boymelgreen2022synthetic}. Studies have also shown that active fluctuations can drive the self-assembly of colloidal systems \cite{mallory2018active}, lead to fluctuation induced phase transitions \cite{di2022evidence}, and enhance the efficiency of microscopic heat engines  \cite{krishnamurthy2016micrometre}. Yet another interesting observation was made in Ref.~\cite{ariga2021noise}, which showed that the molecular motor kinesin \textit{accelerates} in the presence of an active noise in the environment, implying active fluctuations in cells are utilized to promote various physiological processes. This finding was further explored in detail recently in Ref.~\cite{feng2023unraveling}.

Similarly, recent experiments have shown several intriguing phenomena related to the dynamics of an embedded particle as well as swimmers in viscoelastic fluids \cite{shen2011undulatory,camacho2023experimental}.
These observations include the increase in Kramer's rate of transitions \cite{ferrer2021fluid,cherayil2021particle,ginot2022barrier}, circular motion of active particles instead of only persistent random walks \cite{gomez2016dynamics,narinder2019active}, increase in the rotational diffusion coefficient of an active Brownian particle by orders of magnitude compared to that in a viscous fluid \cite{narinder2019active},  and very interestingly, the possibility of attaining the Carnot efficiency in microscopic sterling engine \cite{gomez2021work}. Several recent studies also show that viscoelasticity is essential for facilitating various biological processes \cite{hurst2021intracellular,xie2022contribution,elosegui2023matrix}. 

Compared to the vast literature on the individual fields, the study of the combined effects of active processes in viscoelastic baths is relatively underexplored. Some notable findings in this emerging area include the swelling of a polymer in an active and viscoelastic bath, where the extent of swelling depends on the viscoelastic properties of the environment \cite{vandebroek2015dynamics}, the study of active diffusion of model chromosomal loci driven by athermal noise, revealing the influence of the polymeric effect on the anomalous diffusion exponent \cite{sakaue2017active}, the reduction of quasi-static efficiency in Brownian engines due to the viscoelasticity of the active suspension \cite{guevara2023brownian}, and the implications of complex multi-scale effects leading to time-dependent temperature \cite{ghosh2022active}.

Despite these recent progresses, two interconnected open questions remain: Does the viscoelasticity of the environment affect the performance of micro-engines driven by active processes, and does it influence the arrow of time and fluctuations associated with such processes? In this work, we address these issues through a minimal experiment in a viscoelastic environment. We consider a colloidal particle in a harmonic trap, whose mean position is modulated according to an Ornstein-Uhlenbeck (OU) process. The usage of this type of driving is very common in the context of active matter systems, where the OU process is typically used to mimic the interaction of the passive colloid with the surrounding active particles (such as  molecular motors) \cite{maggi2014generalized,chaki2019effects,guevara2023brownian,paul2022force}. The stochastic thermodynamics of this system in a viscous medium has been extensively studied both experimentally \cite{gomez2010steady,manikandan2021quantitative,dey2022experimental} and theoretically \cite{pal2013work,pal2014work,manikandan2017asymptotics,manikandan2018exact,chaki2018entropy,chaki2019effects}. 

In this paper, we study this model in a homogeneous viscoelastic medium, for which there are no known results to the best of our knowledge. We first analyze the particle dynamics experimentally and theoretically using a generalised Langevin model. The model is solved analytically and the solutions match well with our experimental results. We then proceed to characterizing the non-equilibrium thermodynamics of the system. We calculate the average power supplied to the particle by stochastic driving, which in this case is the same as the thermodynamic \textit{arrow of time} or total entropy production of the process, up to a scaling factor of the temperature of the environment. Interestingly, coupling the system with a viscoelastic medium, as opposed to a viscous medium with the same zero-shear viscosity, leads to approximately $50\%$ enhancement in the mean value of this quantity. This suggests that, rather counter intuitively, viscoelasticity amplifies the process's directionality under identical external driving forces. Furthermore, based on a recent theory for interacting particle systems, we propose that this increase in entropy production in a viscoelastic bath is a generic property, independent of the system's specifics. Next, we look into the fluctuations of the injected power both experimentally and theoretically. In particular, we obtain previously unknown, exact analytical solutions for the  work distribution at short-times, which are non-Gaussian, yet completely determined in terms of the mean and variance of the single-step work fluctuations. The solution obeys a modified fluctuation relation $P(W)/P(-W) = \exp(-2\beta_{ws}\; W)$, where $2\beta_{ws} \ll 1$.  We also obtain an approximate solution for arbitrary time work distribution, which agree excellently with the experimental data. 
Interestingly, we observe a significant reduction in negative work fluctuations, both at short and arbitrary times, in the viscoelastic environment compared to their viscous counterparts. This finding is consistent with the improved arrow of time we observed, highlighting the influence of viscoelasticity on the directionality of the process.

\section{Results}
\begin{figure}[ht]
    \centering
    \includegraphics[width = 0.90\textwidth]{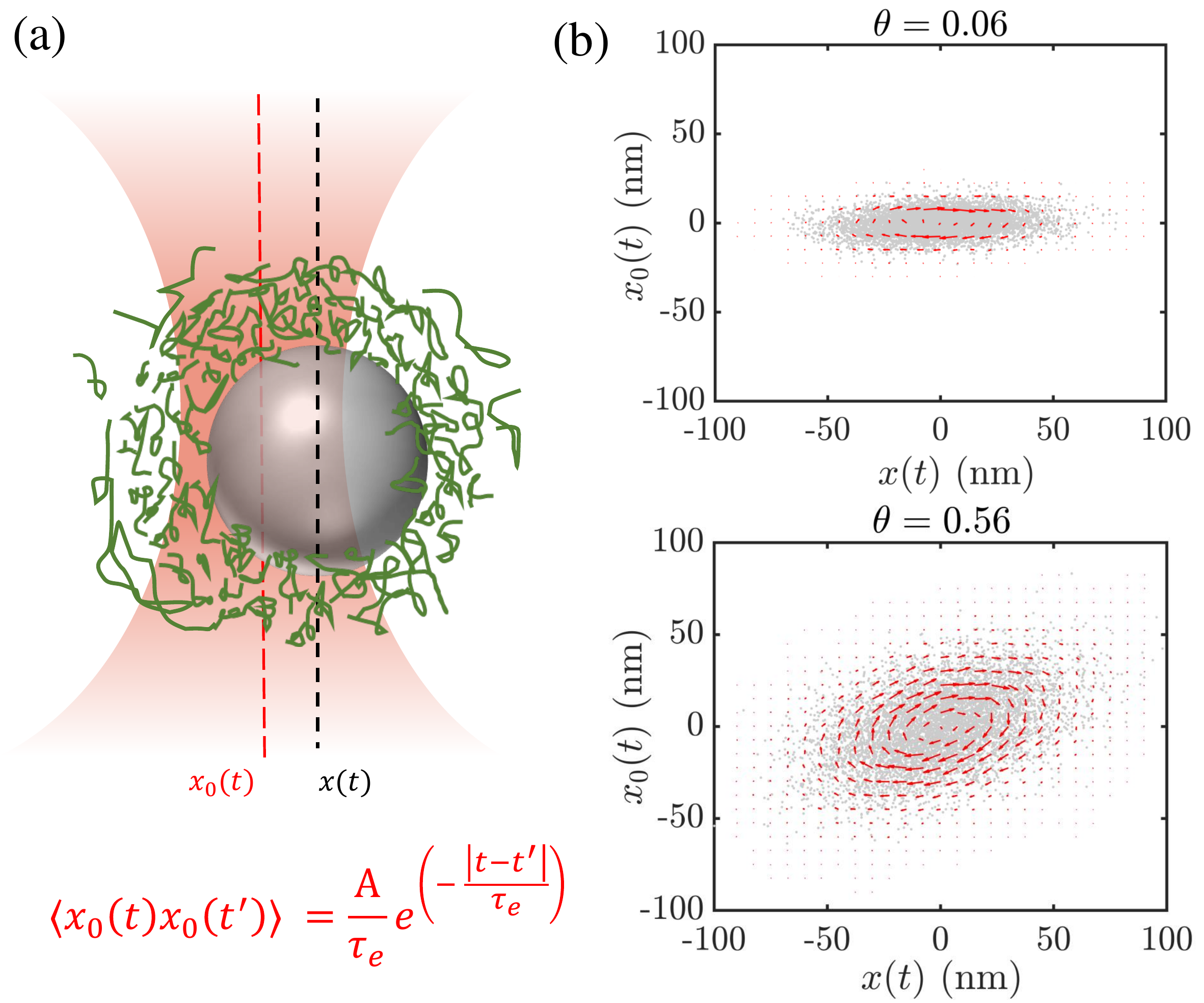}
    \caption{(a) Schematic of a spherical particle trapped in a viscoelastic bath by optical potential whose mean position $({x}_0)$ is externally modulated by Ornstein-Uhlenbeck noise. (b) Typical trajectories of the particle in $({x}, {x}_0)$ plane for different strengths of the external noise (Gray dots) with corresponding non-zero probability flux (Red arrows). }
    \label{fig:diagram}
\end{figure}

Our experimental system primarily consists of an optically trapped, spherical polystyrene particle of diameter $2 a_{0}=3\ \mu m$ (Sigma Aldrich) dispersed in a $0.05\% \ (w/v)$ aqueous solution of poly-ethylene oxide (PEO) [Fig.~\ref{fig:diagram}(a)]. PEO is a polymer that creates a stable configuration in water, which provides elasticity along with its inherent viscosity and renders the solution viscoelastic. The stiffness of the trap is measured to be $12\pm 0.5~  pN/\mu m$. To mimic the active noise in the background of different strengths, we spatially modulate the trapping laser using an acousto-optic modulator (AOM) connected to a signal generator that generates the Orstein-Uhlenbeck (OU) noise of various amplitudes. The trajectory we measure consists of the position of the laser spot $x_0(t)$, as well as the one-dimensional position of the particle $x(t)$ at a spatio-temporal resolution of $1 nm-10kHz$. The readings are obtained using a balanced detector for $\sim 100 s$ (\textit{see  \ref{ap:exp} for details}).

It is known that $x(t)$ is well described using a generalized stochastic differential equation \cite{villamaina2009fluctuation}:
\begin{equation}
m\Ddot{x}(t)=-k[x(t)-x_{0}(t)]-\int_{-\infty}^{t}\Gamma(t-t')\Dot{x}(t')dt'+\eta'(t).
\label{eq1}
\end{equation}
Here the function $\Gamma(t)$ is referred to as the memory kernel. Further, the noise term $\eta'(t)$ satisfied $\langle \eta'(t) \rangle = 0$, and has correlations $\langle \eta'(t)\eta'(t')\rangle=k_{B}T\Gamma(t-t')$; where $k_{B}$ is the Boltzmann constant and $T$ is the temperature of the bath \cite{villamaina2009fluctuation}. 
We further modulate $x_{0}(t)$ according to the Ornstein-Uhlenbeck process:
\begin{equation}
\Dot{x}_{0}(t)=-\frac{x_{0}(t)}{\tau_{e}} +\frac{\sqrt{2A}}{\tau_{e}}\phi_{ext}(t).
\label{eq2}
\end{equation}
Here $\phi_{ext}(t)$ is a zero-mean delta-correlated Gaussian noise with  standard deviation equal to 1, i.e., $\langle \phi_{ext}(t)\rangle = 0$, $\langle \phi_{ext}(t)\phi_{ext}(t^\prime)\rangle = \delta(t-t^\prime)$. $A$ is the strength of that noise (which we experimentally vary using the AOM), and $\tau_{e}$ is the correlation time of that process. The correlation of $x_{0}$ is given by $\langle x_{0}(t)x_{0}(t')\rangle=\frac{A}{\tau_{e}}\exp\left(-\frac{|t-t'|}{\tau_{e}}\right)$.

For the memory kernel $\Gamma$, we consider the Jeffrey's fluid model of viscoelasticity \cite{gomez2015transient,gomez2016dynamics,narinder2019active,ferrer2021fluid,gomez2021work}. According to this model \cite{villamaina2009fluctuation,paul2018free,paul2019active,paul2021bayesian}: 
\begin{equation}
\Gamma(t-t')=2\gamma_{0}\delta(t-t') + \frac{\gamma}{\tau}\exp\left(-\frac{|t-t'|}{\tau}\right).
\label{eq3}
\end{equation}
Here $\tau$ is the relaxation time of the fluid, $\gamma_{0}$ and $\gamma$ are the solvent and the polymer contributions of the frictional coeffecient, respectively \cite{paul2018free}. Notably, the memory kernel will be viscous-like (delta-correlated) in  the $\tau\rightarrow0$ limit as discussed in Refs.\ \cite{ferrer2021fluid,gomez2021work}. Hereafter, we refer to this limit as the viscous limit. 

There are two distinct advantages of using this model for the viscoelastic bath.  Firstly, we can identify the parameters that appear in this model using independent measurements. One such mapping is due to a technique developed in Ref.\ \cite{paul2019single}. In this approach,  one measures the phase response of a particle embedded in the fluid to an external perturbation and computes the relevant parameters by fitting it to theoretical predictions obtained using the Jeffrey's fluid model. By this method, the phase response of the particle over a  wide frequency range can be obtained from a single trajectory using a $3 ~Hz$ square wave as the perturbation (For the details, see \ref{ap:rheology}). We perform this analysis in our setup and the results are shown in Fig.~\ref{fig:rheology_measure} of \ref{ap:rheology}. We obtain the estimates $\tau = 0.018 \pm 0.002 s$ and the ratio between the polymer contribution to the viscosity and solvent contribution to the viscosity, $\mu_{r}(\equiv \gamma/\gamma_0)=1.805\pm0.10$.  
Secondly, as we show in Appendix A\ref{ap:Dynamics}, using the values of parameters inferred in the previous step, and by introducing an additional auxiliary variable $X(t)$, we can write down a Markovian evolution equation in a three dimensional space, whose projection to one dimension results in the Langevin equation in Eq.\ \eqref{eq1}. The same approach has also been used previously in other works to convert a non-Markovian process into Markovian so that the dynamics may be determined conveniently \cite{villamaina2009fluctuation,paul2021bayesian}. Moreover, it is to be noted that this method can also be extended easily to other more general viscoelastic Maxwellian fluids with more than one relaxation time \cite{fricks2009time}.  

Introducing the auxiliary variable $X(t)$, the total dynamical system can then be described using a three-dimensional, linear, stochastic differential equation, 
\begin{equation}
    \boldsymbol{\Dot{Y}}(t)=-\boldsymbol{\lambda}\boldsymbol{Y}(t)+\boldsymbol{D}\boldsymbol{\Phi}(t).
    \label{eq7}
\end{equation}
Where $\boldsymbol{Y}(t) = [x(t),\;X(t),\;x_{0}(t)]^T$, and,
\begin{align}
    \boldsymbol{\lambda}&=\begin{bmatrix}\frac{k}{\gamma_{0}}+\frac{\gamma}{\gamma_{0}\tau} & -\frac{\gamma}{\gamma_{0}\tau} & -\frac{k}{\gamma_{0}}\\-\frac{1}{\tau} & \frac{1}{\tau} & 0\\0 & 0 & \frac{1}{\tau_{e}}\end{bmatrix},\hspace{10mm}
    \boldsymbol{D} =\begin{bmatrix}\sqrt{2D_{0}} & 0 & 0\\0 & \sqrt{2D_{1}} & 0\\0 & 0 & \sqrt{2D_{ext}}\end{bmatrix}.
\end{align}

Here $\bm\Phi(t) = [\phi_{0}(t),\;\phi_{1}(t),\;\phi_{ext}]^T$ is a $3\times1$ dimensional Gaussian-distributed white noise with $\langle \phi_i(t) \rangle = 0$ and $\langle \phi_i(t) \phi_j(s) \rangle = \delta_{ij}\delta(t-s)$. The corresponding diffusion constants are given by $D_{0}=\frac{k_{B}T}{\gamma_{0}}$, $D_{1}=\frac{k_{B}T}{\gamma}$ and $D_{ext}=\frac{A}{\tau_{e}^{2}}$. For the convenience of our discussion, we further define a dimensionless parameter $\theta = \frac{D_{ext}}{D_0} = \frac{A/\tau_e^2}{k_B T/\gamma_0}$.  As we keep all other parameters except $A$ fixed in our experiment, the variation of $\theta$ directly maps to the variation of $A$ (i.e., the strength of the OU noise).

Eq.~\eqref{eq7} is an example of a linear multivariate OU process, and the corresponding Fokker-Planck equation can be solved using standard techniques \cite{risken1996fokker,villamaina2009fluctuation} to compute the steady state distribution $P_{ss}(\bm{Y})$ (\textit{see the Appendix A\ref{ap:Dynamics}}). Due to the presence of the Ornstein-Uhlenbeck driving, this steady state will be not be in equilibrium, being characterized by a nonzero probability flux $\bm{J}_{ss}(\bm{Y})$, as shown in Fig.\ref{fig:diagram}(b). 

\begin{figure}
    \centering
    \includegraphics[width = 0.97\textwidth]{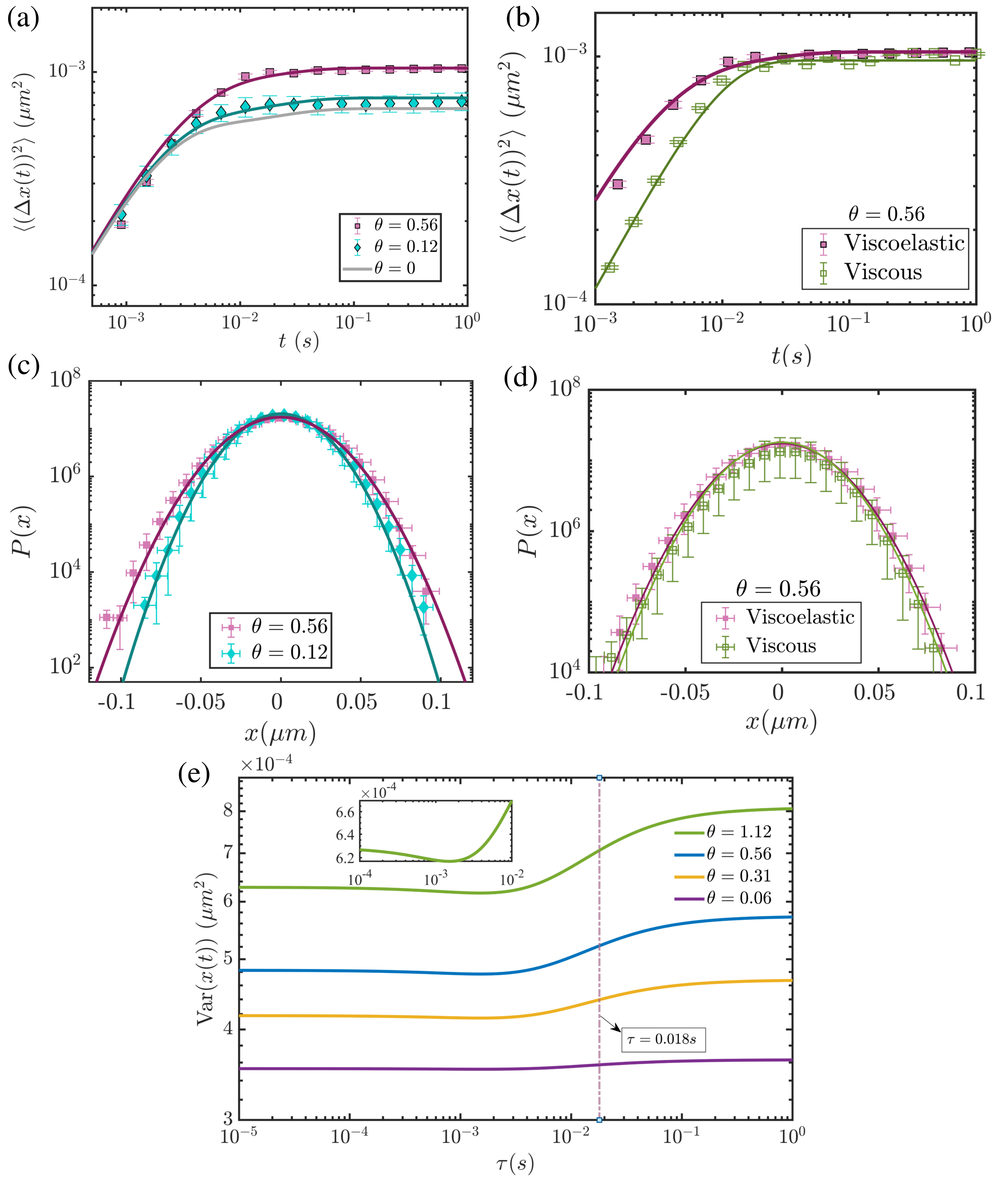}
    \caption{ (a) Mean squared displacement (MSD) of the probe particle in the viscoelastic fluid. (b) Comparison of the MSD of the particle between viscous and viscoelastic fluid for $\theta = 0.56$.  The solid lines denote the theoretical predictions of all the cases. (c) Probability distribution of position fluctuations of the trapped particle in the viscoelastic fluid plotted for different strengths of the external noise. (d) The probability distribution of position fluctuations of the trapped particle in the viscoelastic and in viscous fluid plotted for a fixed strength of the external noise. $\theta = 0.56$. (e) Theoretical estimation of the variance of the displacement of the trapped particle at different noise strengths plotted against the relaxation timescale of the fluid ($\tau$). The dash-dotted line denotes the timescale of the fluid used in the experiment. (Inset) The nonmonotonic behaviour of the variance with $\tau$.}
    \label{fig:ve_dyn}
\end{figure}

Using this analytical solution, we can further obtain the variance of the particle position as,
\begin{equation}
    \text{Var}(x(t))=\frac{k_{B}T}{k} + \frac{Ak}{(\gamma + \gamma_{0} + k\tau)} -\frac{Ak^{2}(\tau_{e}^{2} - \tau^{2})}{(\gamma + \gamma_{0} + k\tau)(\gamma\tau_{e}+\gamma_{0}\tau+\gamma_{0}\tau_{e}+k\tau_{e}^2+k\tau\tau_{e})}.
   \label{eq9}
\end{equation}
Note that we can obtain the variance of $P(x)$ for an active viscous bath (Eq.\eqref{eq:vis_var} of Appendix A\ref{ap:Dynamics}) by considering the limit $\tau\rightarrow0$ in Eq.~\eqref{eq9} which is exactly similar to the related expression given in Ref.\cite{maggi2014generalized}.

Our experimental measurements agree very well with the model. In Fig.~ \ref{fig:ve_dyn}(a), we show the experimentally measured mean-square displacement functions (MSD) along with the corresponding theoretical lines for two different amplitudes of the OU noise. We also show the theoretical MSD line in grey colour for zero external applied noise. Each of these MSD flattens at large times due to the spatial confinement of the particle by optical tweezers, and the corresponding value depends linearly on the effective temperature (defined from the variance of the probability distribution function as $T_{\text{eff}}=k\text{Var}\left(x(t)\right)/k_{B}$). 

In the $\theta \rightarrow 0$ limit, it reaches the equilibrium value $2k_{B}T/k$. In Fig.\ \ref{fig:ve_dyn}(b) we show experimental measures of MSDs in the viscoelastic and viscous fluids for a fixed value of $\theta$ along with corresponding theory curves. For experiments in the viscous fluid, we have chosen a solution of ultra-pure glycerol in water with volume fraction $\sim 0.254$ to reach the viscosity $\sim 2452~\mu$P.s at the room temperature $20^\circ$C, which is very close to the zero-shear viscosity (when $\tau\rightarrow 0$) of the viscoelastic fluid, i.e., $ 2496~\mu\text{P.s}= (\gamma_{0} + \gamma)/6\pi a_{0}$ \cite{cheng2008formula,paul2018two}. Note that due to the change in the refractive index when glycerol is mixed in water, the trap stiffness gets changed. To bring the stiffness back to our fixed value, we changed the trapping laser power appropriately. 

It is clear in Fig.~2(b) that the effective temperature in a viscoelastic fluid is slightly higher as compared to that in a viscous fluid. Further, due to the change in the fluid time scale, a prominent difference in the short-time region is also visible. Fig.~2(c) represents the measured steady-state probability distributions of $x(t)$ for the same two noise amplitudes as in Fig.~2(a) in the viscoelastic fluid. Certainly, due to higher effective temperature for $\theta = 0.56$, the broadening of the corresponding distribution is higher compared to that of $\theta = 0.12$. Interestingly, the higher effective temperature is also visible in Fig.~2(d) which shows the experimental and theoretical steady-state probability distribution of $x(t)$ in viscous and viscoelastic fluids, for a fixed value of $\theta$. 

In Fig. \ref{fig:ve_dyn}(e), we plot $\text{Var}(x(t))$ as a function of $\tau$. We find that for large values of $\tau$, $\text{Var}(x(t))$ saturates to a constant value, which is higher than the corresponding value in the viscous ($\tau \rightarrow 0$) limit. Interestingly, however, $\text{Var}(x(t))$ is non-monotonic in $\tau$ and features a minimum at an intermediate $\tau$ value, that depends on the two other inherent time-scales of the system only and not on $\theta$ (see Eq.\eqref{eq:var_min_tau} of the Appendix A\ref{ap:Dynamics} for the expression of $\tau$ corresponding to the minima). We show this in the inset of   Fig.\ \ref{fig:ve_dyn}(e). Note that this is purely an effect of viscoelasticity of the medium since the effect of the OU driving alone does not result in this non-monotonic behaviour (see Ref.\ \cite{maggi2014generalized}).

\begin{figure}
     \centering      
     \includegraphics[width = 0.90\textwidth]{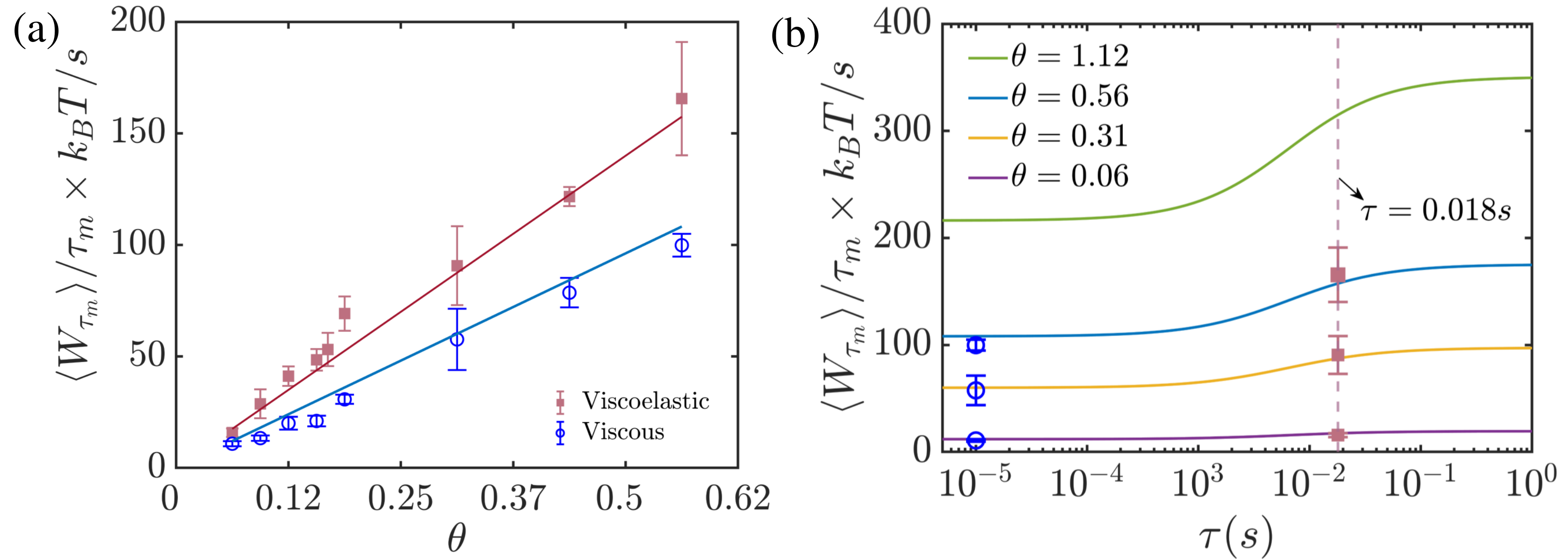}
     \caption{(a) Rate of mean work done by the external modulation on the trapped Brownian particle in viscoelastic bath and viscous bath ( $\tau\rightarrow0$)  plotted against the strength of the external noise. The solid lines denote the theoretical predictions. (b) Variation of work with the relaxation timescale of the medium plotted for different strengths of the external noise along with experimental points. The dash-dotted line denotes the timescale of the fluid used in the experiment. }
     \label{fig:meanwr_A}
 \end{figure}

Now we move on to study the non-equilibrium thermodynamics of this system. To this end, we look at the statistics of the injected mechanical work done by the external forcing on the colloidal particle, defined as \cite{sekimoto1998langevin,seifert2012stochastic,pal2014work,manikandan2017asymptotics,dey2022experimental}:
\begin{equation}
    W_{\tau_{m}}=\frac{k}{k_{B}T}\int_{t}^{t+\tau_{m}}\Dot{x}(t')x_{0}(t')dt'.
    \label{eq11}
\end{equation}
Using this expression, the average power delivered to the particle can be straightforwardly computed from the experimental data both in viscous and viscoelastic fluids. The results are shown in Fig.~\ref{fig:meanwr_A}(a), as a function of noise-strength $\theta$. We find that $\langle W_{\tau_{m}} \rangle$ is positive as expected from the Second Law, and increases linearly as a function of $\theta$. The estimates also agree well with the analytically computed estimate (solid lines in Fig.~\ref{fig:meanwr_A}(a), \textit{see Appendix A\ref{ap:work_fluctuation} for details}) from the model, given by,
\begin{equation}
  \langle W_{\tau_{m}}\rangle=\frac{k\tau_{m}}{k_{B}T}\frac{Ak(\tau + \tau_{e})}{\tau_{e}(\gamma_{0}\tau + \gamma\tau_{e} + \gamma_{0}\tau_{e} + k\tau_{e}^{2} + k\tau\tau_{e})}.
  \label{eq12}
\end{equation}
Interestingly, the obtained average power in a viscoelastic bath is significantly higher compared to the same for a `memory-less' Newtonian bath with our chosen parameters. This implies that for the same amplitude of external driving, a higher power is delivered to a particle in a viscoelastic bath compared to its viscous counterpart (See the Eq.\eqref{eq:ve_meanwr_th} and Eq.\eqref{eq:vis_meanwr_th} of Appendix A\ref{ap:work_fluctuation} for corresponding expressions in terms of $\theta$ both for viscous and viscoelastic fluids for a comparison). 
 
Even though our finding is for a particular example of a non-equilibrium system, it is possible to argue that this is a generically expected property for any non-equilibrium system  in a viscoelastic environment.  The argument is based on the observation that the average injected power for this system is the same as the average total entropy production, irrespective of the rheological properties of the bath (\textit{See  \ref{ap:EPR}}). This is because the injected power is the only time extensive source of dissipation in the system. Furthermore, for systems in contact with a heat reservoir, this quantity can be considered as representative of the global arrow of time, defined as, 
\begin{equation}
    \text{Arrow of time} =\frac{1}{T}\sum_{\bm{Y}(\cdot)} P[\bm{Y}(\cdot)]\log\frac{P[\bm{Y}(\cdot)]}{P[\tilde{\bm{Y}}(\cdot)]},
\end{equation}
where $\bm{Y}(\cdot) = \lbrace \bm{Y}(t) \rbrace _{t = 0}^T$ is the trajectory of the system measured for a  time duration $T$, and $\tilde{\bm{Y}}(t)$ is its time reversed counterpart defined as $\Tilde{\bm{Y}}(t) = \bm{Y}(T-t)$.
It was recently shown that the presence of interactions increases the evidence of the global arrow of time \cite{lynn2022decomposing}. In a viscoelastic bath, we know already that the polymers in the viscoelastic solution contribute to additional interactions and that can significantly affect the statistics of the position of the particle. Thus we argue that the increase in entropy production is due to the interaction of the colloidal particle with polymers in the solution. We note that a similar argument can be put forward on the basis of the theoretical model, where the random variable $X(t)$ is the additional interacting degree of freedom which is absent in a viscous bath. 

A natural question that then arises is whether the arrow of time of the process can be continuously increased by working with a medium of increased viscoelastic response. Although this is experimentally challenging to address, we can seek an answer using the expressions derived using the theoretical model.  In Fig.~\ref{fig:meanwr_A}(b), we show that in the limit of increasing relaxation time $\tau$ of the viscoelastic bath, $\langle W_{\tau_m} \rangle$ increases and saturates to a constant value that depends only on the properties of the trap parameters. However, in contrast to the non-monotonic behavior observed for $\text{Var}(x(t))$, we find that $\langle W_{\tau_m} \rangle$ exhibits a monotonically increasing function of $\tau$. This observation reinforces the suggestion that hidden interactions play a significant role in the amplification of the thermodynamic arrow of time. 

Finally, we look at the fluctuations of the injected power by computing the probability distribution $P(W_{\tau_m})$. In certain cases, it is known that this probability distribution obeys a fluctuation theorem of the form:
\begin{equation}
    \ln \frac{P(W_{\tau_m} = w)}{P(W_{\tau_m} = -w)} \rightarrow \frac{w}{k_B T},\ \ \ \tau_m >>\tau_l,
    \label{eq:wft}
\end{equation}
where $\tau_l$ denotes the largest time scale of the system and $k_B$ is the Boltzmann constant. The work fluctuation theorem has already been tested and validated in diverse experiments and numerical simulations. Similar setups to ours include an optically trapped particle dragged with a constant velocity \cite{wang2005demonstration}, Brownian oscillator with deterministic periodic driving in the non-markovian bath\cite{mai2007nonequilibrium}, a colloidal particle trapped in a bi-stable potential subjected to periodic modulation \cite{jop2008work}, Brownian particle trapped in an optical trap with discontinuously varying trap stiffness \cite{carberry2004fluctuations,carberry2007optical} and also in other cases such as electronic circuits \cite{garnier2005nonequilibrium}, mechanical oscillators \cite{joubaud2007fluctuation}, etc. However, violations are also found in many systems, e.g., a colloidal particle under the action of a Gaussian distributed white and coloured noise \cite{farago2002injected,farago2004power,gomez2010steady,chaki2018entropy}, wave turbulence \cite{falcon2008fluctuations}, an RC circuit driven with a stochastic voltage \cite{falcon2009fluctuations}, a rotating active Brownian particle in a viscoelastic medium \cite{narinder2021work}, etc. These violations have been explored very well, and it is known that the primary reason for the violation is the presence of time intensive contributions to the total entropy production, which can significantly contribute to fluctuations at short times and in continuous state space systems \cite{verley2014work,manikandan2019efficiency}.

An exact expression for the work probability distribution that appears in the fluctuation relations is known only in a handful of cases. For example, in the case of a colloidal particle in a harmonic trap with the mean position externally controlled, for any arbitrary deterministic driving, the work distribution can be shown to be a Gaussian \cite{Vanzone1,Nickelsen2011}. The case where the stiffness is time dependently controlled is one of the simplest models showing non-Gaussian work fluctuations \cite{speck}. Another system for which an exact calculation of the work distribution has been carried out is the colloidal particle in a logarithmic-harmonic potential \cite{loghar}, which also has non-Gaussian work fluctuations. A similar occurrence of non-Gaussian work fluctuations in a complex fluid in the context of an optically trapped particle with deterministically varying trap stiffness has also been discussed by \textit{Carberry et.al.} in Ref.\cite{carberry2007optical}. In \cite{Saha2015}, exact work statistics have also been obtained for a colloidal particle in the uniformly dragged two dimensional confining potential, and in the presence of additional non-conservative forces such as torques. In this case, however, the work distribution is found to be a Gaussian, although it deviates from the conventional  work fluctuation theorem. A general framework for computing the large-time form of the work distribution, for linear Langevin processes, is described in Refs. \cite{mai2007nonequilibrium,pal2013work,pal2014work}.  The method is based on the calculation of the corresponding moment-generating function using a large deviation technique and gives closed form expressions exact to sub leading order in $1/t$. The methods developed in Refs. \cite{manikandan2017asymptotics,manikandan2018exact} provides exact analytical expressions for the moment generating function for any value of $t$ for the same class of systems. However, inverting them to obtain a closed form expression for $P(W,t)$ is challenging.

\begin{figure}
     \centering
      \includegraphics[width = 0.95\textwidth]{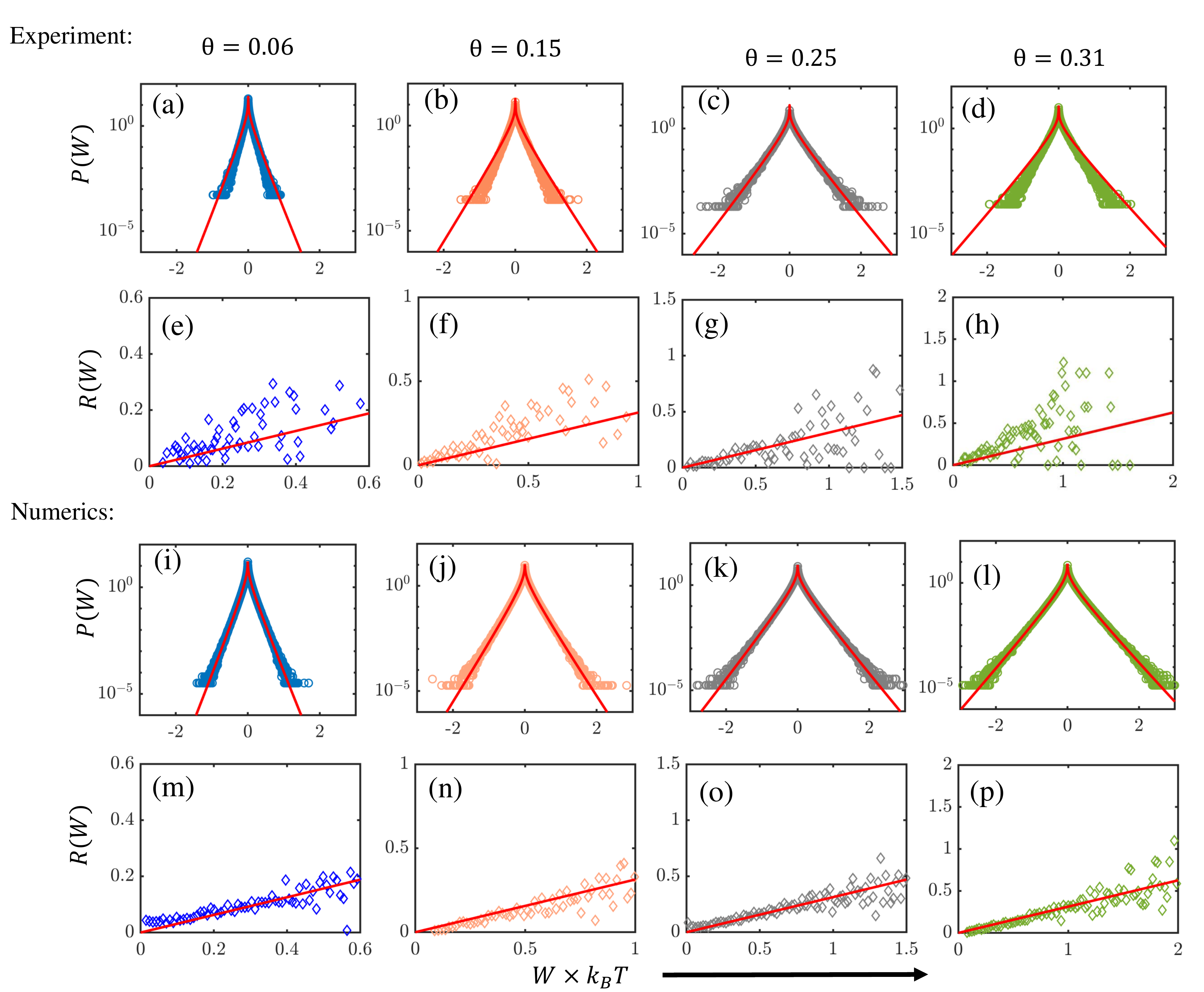}
     \caption{(a)-(d) Probability density function of single-step work done (calculated from experimental data) by the external modulation on the trapped particle in the viscoelastic bath plotted for different strengths of the external noise. Corresponding probability ratios are plotted in (e)-(h). (i)-(l) The numerically calculated probability density functions of single-step work done are shown for different strengths of the modulation. (m)-(n) Numerically estimated probability ratios for the corresponding cases.  Theoretical estimations in all the cases are represented by solid red lines. }
     \label{fig:single_step_pdfw}
 \end{figure}

 \begin{figure}
     \centering
      \includegraphics[width = 0.8\textwidth]{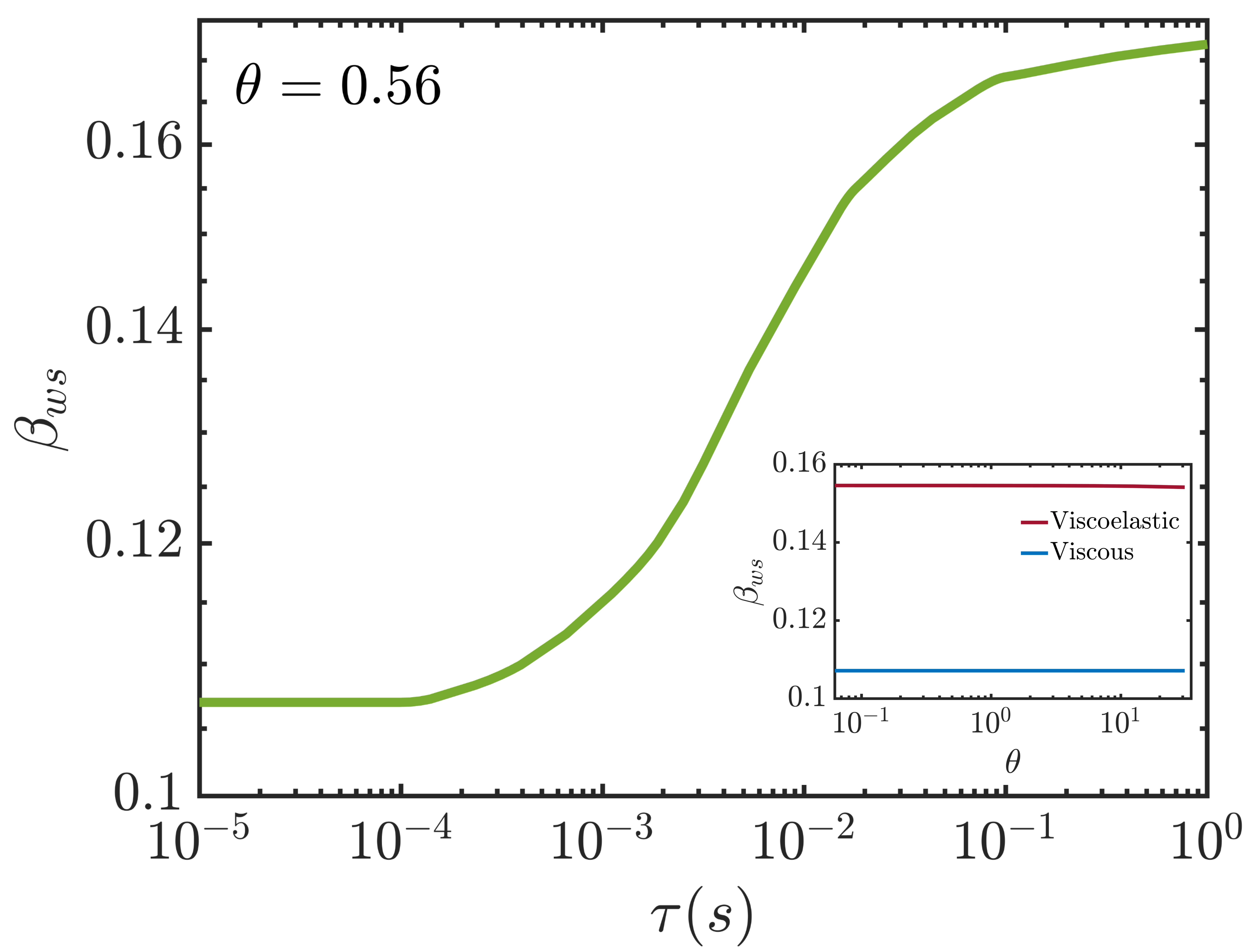}
     \caption{ Variation of $\beta_{ws}$ with $\tau$. {(Inset) $\beta_{ws}$ shows extremely slow variation with $\theta$ for both viscous and viscoelastic baths.}}
     \label{fig:pdfw_A}
 \end{figure}
 
Here, using a novel technique, we are able to compute a closed form expression of the probability distribution $P(W_{\tau_m})$ in the single-step (short-time) limit. This limit has received a lot of attention recently, primarily in the context of saturation of the thermodynamic uncertainty relation \cite{manikandan2020inferring,otsubo2020estimating,van2020entropy,manikandan2021quantitative, otsubo2022estimating,manikandan2022nonmonotonic,kumar2022inference}. In particular, it is known that in the strict limit of $\tau_m \rightarrow 0$ fluctuations of any time-integrated current in an overdamped diffusive system is Gaussian. The single-step limit we consider here is the closest to the short-time limit (that is experimentally accessible),  where the distribution is expected to be non-Gaussian.  The technique we use is based on the fact that the single-step work $W$ appears as the product of two zero-mean Gaussian distributed quantities $\Dot{x}(t)$ and $x_{0}$(t) which are cross-correlated (see Eq.~\eqref{eq11}). The probability distribution of $W$, therefore, can be obtained directly from the joint probability distribution function of $\Dot{x}(t)$ and $x_{0}$(t) as a skewed zeroth-order modified Bessel function \cite{cui2016exact}
\begin{equation}
P(W)=\mathcal{A}_{ws}\exp{(\beta_{ws} W)}\text{K}_{0}(\alpha_{ws}|W|) 
\label{short-time_pdf},
\end{equation}
where
\begin{align}
\begin{split}
\beta_{ws}&=\frac{\langle\text{W}\rangle}{\text{Var} (\text{W}) - 2\langle\text{W}\rangle^{2}},\\
\alpha_{ws}&=\left[\beta_{ws}^{2} + \frac{\beta_{ws}}{\langle \text{W}\rangle}\right]^{\frac{1}{2}},\\
\mathcal{A}_{ws}&=\frac{\sqrt{\alpha_{ws}^{2} - \beta_{ws}^{2}}}{\pi}.
\label{eq15}
\end{split}
\end{align}
The subscript `$ws$' stands for single-step work. See Appendix A\ref{ap:short_time_wf} for a detailed derivation of the expression above. 

Clearly, the expression above is non-Gaussian. However, interestingly, it is entirely determined in terms of the mean and variance of single-step work. Further note that, except for an exponential factor of $e^{\beta_{ws} \text{W}}$, the distribution $P(W)$ does not depend on the sign of $W$. Hence, it is straightforward to write down a modified fluctuation relation, of the form:
\begin{equation}
R(W) = \ln{\frac{P(W)}{P(-W)}}=2\beta_{ws}W.
\label{eq:FT}
\end{equation}
In Fig.\ \ref{fig:single_step_pdfw} (a) -- (d) and (i) -- (l), we show the excellent agreement between this analytically computed expression with the experimental as well as numerical data. 
In Fig.\ \ref{fig:single_step_pdfw} (e) -- (h), and (m) -- (p) we again verify this fluctuation theorem using both experimental and numerical data. 

By definition, $\beta_{ws}$ corresponds to the relative likelihood of having a positive work fluctuation over the negative one of the same magnitude. It is hence interesting to check how this factor is affected by the viscoelasticity of the medium. In Fig.\ \ref{fig:pdfw_A}, we plot $\beta_{ws}$ as a function of the viscoelastic fluid time-scale $\tau$ and find that it monotonically increases with it. This trend is qualitatively similar to that observed for the average entropy production rate (see Fig.\ \ref{fig:meanwr_A}b), and shows that for a fixed strength of non-equilibrium driving, a positive work fluctuation is relatively more favoured in a viscoelastic bath with a higher value of $\tau$. We have further found that the effect of viscoelasticity on $\beta_{ws}$ is much more significant as compared to the effect of $\theta$ on the same. In the inset of Fig.\ \ref{fig:pdfw_A}, we analytically compare $\beta_{ws}$ obtained for a range of $\theta$ values for the viscoelastic case and viscous case. We find that $\beta_{ws}$ for the viscoelastic case is always higher than $\beta_{ws}$ for the viscous case, but in both cases, it varies negligibly with $\theta$. 
\begin{figure}
    \centering
    \includegraphics[width = 0.90\textwidth]{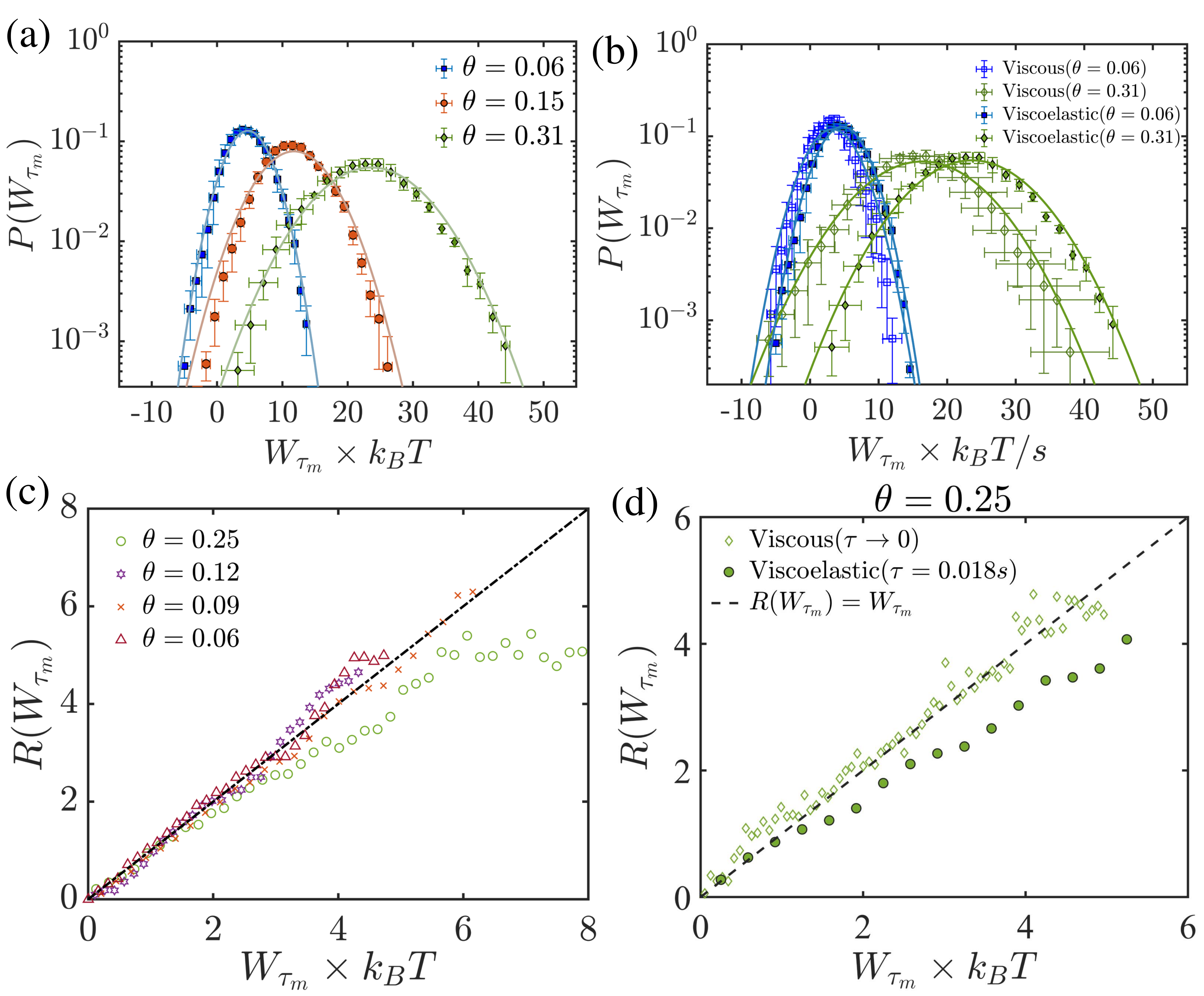}
    \caption{(a) Probability density function of work done (calculated from experimental data) by the external modulation on the trapped particle in the viscoelastic bath plotted for different strengths of the external noise. Theoretical estimations in all the cases are represented by solid lines. 
    (b) Probability of positive fluctuations of work are enhanced for the viscoelastic bath as seen from the comparison of the same for the viscous bath. Theoretical estimations (Eq.\eqref{eq:co_wpdf_l} of Appendix \ref{ap:cumulative_wf}) in all the cases are represented by solid lines.
    (c) $R(W_{\tau_{m}})$ (estimated from the experimental data) plotted as a function of $W_{\tau_{m}}$ for different strengths of the external modulation when the particle is in the viscoelastic environment. 
    (d) $R(W_{\tau_{m}})$ for the viscoelastic bath is found to be deviating from the usual work fluctuation theorem (\textit{black dashed line}) with smaller external noise than the same for the viscous bath.}
    \label{fig:pdf_wft_ve_vis}
\end{figure}

Next, we consider finite time work distributions, which represent the distributions of cumulative work $W_{\tau_{m}}$ obtained using Eq.~\eqref{eq11}. To approximate this distribution, we assume an uncorrelated time series of short-time work and utilize the probability distribution function from Eq.\ \eqref{short-time_pdf}. However, in reality, the short-time work time series is not perfectly uncorrelated, rendering the expression invalid. Nevertheless, we find that by modifying our approximations based on the mean and variance of $W_{\tau_{m}}$ directly calculated from the Langevin dynamics of the process, we can achieve good agreement with experimentally and numerically measured distributions, particularly for typical work fluctuations (see Eq.\eqref{eq:co_wpdf_l} of Appendix A\ref{ap:cumulative_wf}). This is illustrated in Fig.~\ref{fig:pdf_wft_ve_vis}(a) and (b). We also observe that positive work fluctuations are significantly more favoured in the viscoelastic case as compared to the corresponding viscous limit. In Fig.\ \ref{fig:pdf_wft_ve_vis} (c) and (d), we further compute the ratio of probabilities [$R(W_{\tau_m})$] for positive and negative work fluctuations. While the ratio is not linear for all work values, it shows linear behaviour with a unit slope for small work values, deviating for larger ones. The deviations are greater in a viscoelastic bath compared to a viscous environment, which again substantiates stronger deviations from equilibrium.

\section{Conclusions}
In summary, we have shown that the viscoelasticity of the environment has a significant impact on the thermodynamics and fluctuations of a driven colloidal particle. Specifically, we have found a substantial increase (around $50\%$) in the mean injected power to the system when compared to the same system in a viscous bath under the same external driving strength, even though the steady-state variance of the position fluctuations gets nominally (around $10\%$) enhanced. Additionally, we have observed lower negative work fluctuations across a wide range of driving amplitudes. Since the injected power corresponds to the thermodynamic arrow of time (entropy production) in this process, these findings suggest an enhanced directionality in a driven process when it takes place in a viscoelastic bath rather than a viscous bath with the same zero-shear viscosity.

In light of recent work \cite{lynn2022decomposing}, our findings suggest that the influence of viscoelasticity on directionality and thermodynamics may extend beyond the specific experimental setup studied in this work, potentially affecting a wide range of driven processes in biological contexts. For example, it will be interesting to study whether the viscoelasticity of the medium affects the directionality of molecular motors studied \textit{in vitro}. Our results also suggest that viscoelastic environments may provide a more controlled and predictable environment for driven processes, potentially leading to improved performance and reliability in practical applications, such as the assembly of nanoscale materials \cite{liz2003nanoscale}. We plan to address some of these issues in a future work. 

\section*{Acknowledgments}
The work was supported by IISER Kolkata, an autonomous teaching and research institute supported by the Ministry of Education, Government of India, and the Science and Engineering Research Board, Department of Science and Technology, Government of India, through the research grant CRG/2022/002417. BD is thankful to the Ministry Of Education of
Government of India for financial support through the Prime Minister’s Research Fellowship (PMRF) grant. SP is supported financially by the postdoctoral research fellowship of the IISER Kolkata. SKM acknowledges the Knut and Alice Wallenberg Foundation for financial support through Grant No. KAW 2021.0328.

%\section*{Author Declarations}
\section*{Conflict of Interest}
The authors have no conflicts to disclose.
\section*{Data Availability Statement}

The data that support the findings of this study are available from the corresponding author upon reasonable request.

\appendix
\renewcommand{\thesubsection}{\Roman{subsection}}
\section{ Analytical way to calculate PDF of stochastic work by external force  }
\label{ap:ana_cal}
\subsection{Dynamics of the system}
\label{ap:Dynamics}
We model the one-dimensional dynamics of an externally forced spherical Brownian particle of mass $m$ confined in a harmonic potential of stiffness $k$ in a homogeneous and isotropic viscoelastic medium with the following stochastic differential equation \cite{villamaina2009fluctuation}:
\begin{equation}
m\Ddot{x}(t)=-k[x(t)-x_{0}(t)]-\int_{-\infty}^{t}\Gamma(t-t')\Dot{x}(t')dt'+\eta'(t)
\label{eq1a}
\end{equation}
where $x_{0}(t)$ is the random position of the potential minimum at time $t$ originated from the equation
\begin{equation}
\Dot{x}_{0}(t)=-\frac{x_{0}(t)}{\tau_{e}} +\frac{\sqrt{2A}}{\tau_{e}}\phi_{ext}(t).
\label{eq2a}
\end{equation}
$\phi_{ext}(t)$ is a zero-mean delta-correlated Gaussian noise with unit standard deviation, and $A$ is the strength of that noise. Note that $x_{0}(t)$ is an Ornstein-Uhlenbeck (OU) process with correlation time $\tau_{e}$, with the correlation given by $\langle x_{0}(t)x_{0}(t')\rangle=\frac{A}{\tau_{e}}\exp\left(-\frac{|t-t'|}{\tau_{e}}\right)$. $\eta'(t)$ is the Gaussian-distributed stochastic noise representing the thermal agitation of the fluid molecules. To satisfy the \textit{fluctuation-dissipation theorem} (FDT) in equilibrium, the correlation of the noise is related to the memory kernel $\Gamma(t)$ as $\langle \eta'(t)\eta'(t')\rangle=k_{B}T\Gamma(t-t')$, where $k_{B}$ is the Boltzmann constant and $T$ is the temperature of the bath. As a viscoelastic model, we consider Jeffrey's fluid model where the memory Kernel is of the kind: 
\begin{equation}
\Gamma(t-t')=2\gamma_{0}\delta(t-t') + \frac{\gamma}{\tau}\exp\left(-\frac{t-t'}{\tau}\right),
\label{eq3a}
\end{equation}
where $\tau$ is the relaxation time of the fluid; $\gamma_{0}$ and $\gamma$ are the solvent and the polymer contributions respectively. The thermal noise $\eta'(t)$ can be represented as the sum of two zero-mean independent Gaussian processes as $\eta'(t)=\eta'_{0}(t)+\eta'_{1}(t)$, where the correlations of these two processes are given by $\langle \eta'_{0}(t)\eta'_{0}(t')\rangle=2k_{B}T\gamma_{0}\delta(t-t')$ and $\langle \eta'_{1}(t)\eta'_{1}(t')\rangle=k_{B}T\frac{\gamma}{\tau}\exp\left(-\frac{|t-t'|}{\tau}\right)$ respectively with $\gamma_0 = 6\pi a_0 \mu_s$ and $\gamma = 6\pi a_0 \mu_p$. $\mu_p$ and $\mu_s$ are the viscosities of the polymer and the solvent, respectively. At the overdamped limit (where the inertial effect is negligible), after carrying out an integration by parts, Eq.~\eqref{eq1a} can be written as
\begin{equation}
\Dot{x}(t)= -\left(\frac{k}{\gamma_{0}}+\frac{\gamma}{\gamma_{0}\tau}\right)x(t)+\frac{\gamma}{\gamma_{0}\tau^{2}}\int_{-\infty}^{t}\exp\left(-\frac{t-t'}{\tau}\right)x(t')dt'+\eta_{0}(t) + \eta_{1}(t) + \frac{k}{\gamma_{0}}x_{0}(t).
\label{eq4a}
\end{equation}
We define $\eta_{0}(t)=\eta'_{0}(t)/\gamma_{0}$ and $\eta_{1}(t)=\eta'_{1}(t)/\gamma_{0}$. Now, in order to model the system as Markovian we introduce an auxiliary variable
\begin{equation}
X(t)=\frac{1}{\tau}\int_{-\infty}^{t}\exp\left(-\frac{t-t'}{\tau}\right)\left[x(t')+\tau\sqrt{\frac{2k_{B}T}{\gamma}}\phi_{1}(t')\right]dt'
\label{eq5a}
\end{equation}
and combining Eqns.~\eqref{eq2a}, \eqref{eq4a} and \eqref{eq5a} we can write in the matrix form,
\begin{equation}
\begin{bmatrix}\Dot{x}(t)\\\Dot{X}(t)\\\Dot{x}_{0}(t)\end{bmatrix}=-\begin{bmatrix}\frac{k}{\gamma_{0}}+\frac{\gamma}{\gamma_{0}\tau} & -\frac{\gamma}{\gamma_{0}\tau} & -\frac{k}{\gamma_{0}}\\-\frac{1}{\tau} & \frac{1}{\tau} & 0\\0 & 0 & \frac{1}{\tau_{e}}\end{bmatrix}\begin{bmatrix}x(t)\\X(t)\\x_{0}(t)\end{bmatrix} + \begin{bmatrix}\sqrt{2D_{0}} & 0 & 0\\0 & \sqrt{2D_{1}} & 0\\0 & 0 & \sqrt{2D_{ext}}\end{bmatrix}\begin{bmatrix}\phi_{0}(t)\\\phi_{1}(t)\\\phi_{ext}(t)\end{bmatrix}
\label{eq6a}
\end{equation}
where $\phi_{0}$, $\phi_{1}$ and $\phi_{ext}$ are Gaussian-distributed white noises with mean zero and unit standard deviation. $D_{0}=\frac{k_{B}T}{\gamma_{0}}$, $D_{1}=\frac{k_{B}T}{\gamma}$ and $D_{ext}=\frac{A}{\tau_{e}^{2}}$. Considering
\begin{gather*}
    \boldsymbol{Y}(t)=\begin{bmatrix}x(t)\\X(t)\\x_{0}(t)\end{bmatrix},
    \hspace{1mm}\boldsymbol{\lambda}=\begin{bmatrix}\frac{k}{\gamma_{0}}+\frac{\gamma}{\gamma_{0}\tau} & -\frac{\gamma}{\gamma_{0}\tau} & -\frac{k}{\gamma_{0}}\\-\frac{1}{\tau} & \frac{1}{\tau} & 0\\0 & 0 & \frac{1}{\tau_{e}}\end{bmatrix},\\
    \hspace{1mm}\boldsymbol{D}=\begin{bmatrix}\sqrt{2D_{0}} & 0 & 0\\0 & \sqrt{2D_{1}} & 0\\0 & 0 & \sqrt{2D_{ext}}\end{bmatrix},
    \hspace{1mm}\boldsymbol{\Phi}(t)=\begin{bmatrix}\phi_{0}(t)\\\phi_{1}(t)\\\phi_{ext}(t)\end{bmatrix}\nonumber
\end{gather*}
we can rewrite the Eq.~\eqref{eq6a} as the following:
\begin{equation}
    \boldsymbol{\Dot{Y}}(t)=-\boldsymbol{\lambda}\boldsymbol{Y}(t)+\boldsymbol{D}\boldsymbol{\Phi}(t).
    \label{eq7a}
\end{equation}
Clearly, Eq.~\eqref{eq7a} is a multivariate Ornstein-Uhlenbeck process with correlation matrix \cite{villamaina2009fluctuation}
\begin{equation}
    \boldsymbol{C}=\exp(-\boldsymbol{\lambda}t)\boldsymbol{\sigma}.
    \label{eq8a}
\end{equation}
$\boldsymbol{\sigma}$ is the covariance matrix determined by the stationary condition of the process in terms of \textit{Lyapunov equation} $\boldsymbol{\lambda}\boldsymbol{\sigma}+(\boldsymbol{\lambda}\boldsymbol{\sigma})^{T}=\boldsymbol{D}\boldsymbol{D}^{T}$ \cite{risken1996fokker}. The corresponding steady-state probability distribution ($P_{ss}(\mathbf{Y})$) and the corresponding current ($\mathbf{J}_{ss}(\mathbf{Y})$) can be described as:
\begin{align}
\begin{split}
    P_{ss}(\mathbf{Y}) = (2\pi \sqrt{\det \boldsymbol{ \sigma}})^{-1} e^{-\frac{1}{2}\mathbf{Y}^T\boldsymbol{\sigma}^{-1}\mathbf{Y}} \\
    \mathbf{J}_{ss}(\mathbf{Y}) =  (-\boldsymbol{\lambda}\mathbf{Y} + \mathbf{D}_{f}\boldsymbol{\sigma}^{-1}\mathbf{Y})P_{ss}(\mathbf{Y}),
    \end{split}
    \label{eq:prob_current1}
\end{align}
where $\mathbf{D}_f = \frac{1}{2}\mathbf{D}\mathbf{D}^T$.
The variance of $x(t)$ is given by $\sigma_{11}$ and after solving the stationary condition, we get the following expression:
\begin{equation}
    \text{Var}(x)=\sigma_{11}=\frac{k_{B}T}{k} + \frac{Ak}{(\gamma + \gamma_{0} + k\tau)}-\frac{Ak^{2}(\tau_{e}^{2} - \tau^{2})}{(\gamma + \gamma_{0} + k\tau)(\gamma\tau_{e}+\gamma_{0}\tau+\gamma_{0}\tau_{e}+k\tau_{e}^2+k\tau\tau_{e})}.
    \label{eq9a}
\end{equation}
Interestingly, $\text{Var}(x)$ shows  non-monotonic dependence (Fig.\ref{fig:ve_dyn}(e) of main text)  on the relaxation timescale of the fluid ($\tau$)  with the minima at $\tau = \tau_{min}$ such that, 
\begin{equation}
    \label{eq:var_min_tau}
    \tau_{min} = \tau_e\ \frac{[(\gamma_0 + \gamma)(2\gamma_0+\gamma+2k\tau_e)]^{1/2}-(\gamma_0+\gamma)}{\gamma_0+2k\tau_e}
\end{equation}
It is clear that $\tau_{min}$ has a non-trivial dependence on the other timescales of the system and is independent of the strength of external perturbation ($A$). \\

The expression for the variance of $P(x)$ for a Newtonian fluid can be easily calculated by considering the limit $\tau\rightarrow{0}$ in $\sigma_{11}$, which becomes
\begin{equation}
    \sigma_{N}=\frac{k_{B}T}{k}+\frac{Ak}{\gamma + \gamma_{0}+k\tau_{e}}.
    \label{eq:vis_var}
\end{equation}
Note that this expression is similar to that obtained in an active viscous bath given in Ref. \cite{maggi2014generalized}.
The position auto-correlation function ($\text{ACF}_{x}$) of the particle is $C_{11}(t)$, i.e., $\langle x(t)x(t')\rangle=C_{11}(t-t')$, that can be easily calculated from Eq.~\eqref{eq8a}. The mean-square displacement (MSD) of the particle for our system is related to the $\text{ACF}_{x}$ as $\langle(x(t)-x(0))^{2}\rangle=2[C_{11}(0)-C_{11}(t)]$ \cite{paul2018free}.

\subsection{Work fluctuations} 
\label{ap:work_fluctuation}
The stochastic mechanical work by the external force over a time $\tau_{m}$ is defined as the following \cite{sekimoto1998langevin,seifert2012stochastic,pal2014work,manikandan2017asymptotics,dey2022experimental}:
\begin{equation}
    W_{\tau_{m}}=\frac{k}{k_{B}T}\int_{t}^{t+\tau_{m}}\Dot{x}(t')x_{0}(t')dt',
    \label{eq10a}
\end{equation}
and the mean of work is given by,
\begin{equation}
    \langle W_{\tau_{m}}\rangle=\frac{k}{k_{B}T}\int_{t}^{t+\tau_{m}}\langle \Dot{x}(t')x_{0}(t')\rangle dt'.
    \label{eq11a}
\end{equation}
Considering $c=\begin{bmatrix}x_{0}(t) & 0 & 0\end{bmatrix}$ and multiplying with Eq.~\eqref{eq7a} from the left we get
$$\langle\Dot{x}(t)x_{0}(t)\rangle=-\left(\lambda_{11}\langle x(t)x_{0}(t)\rangle + \lambda_{12}\langle X(t)x_{0}(t)\rangle + \lambda_{13}\langle x_{0}(t)^{2}\rangle\right) + 2D_{0}\langle x_{0}(t)\phi_{0}(t)\rangle.$$
As, $\langle x_{0}(t)\phi_{0}(t)\rangle=0$ and $\langle x(t)x_{0}(t)\rangle=\sigma_{13}$, $\langle X(t)x_{0}(t)\rangle=\sigma_{23}$, $\langle x_{0}(t)^{2}\rangle=\sigma_{33}$; we finally have,
\begin{eqnarray}
 \langle\Dot{x}(t)x_{0}(t)\rangle &=& -\left(\lambda_{11}\sigma_{13} + \lambda_{12}\sigma_{23} + \lambda_{13}\sigma_{33} \right)\nonumber\\
 &=&\frac{Ak(\tau + \tau_{e})}{\tau_{e}(\gamma_{0}\tau + \gamma\tau_{e} + \gamma_{0}\tau_{e} + k\tau_{e}^{2} + k\tau\tau_{e})}
 \label{eq12a}
\end{eqnarray}
Therefore,
\begin{equation}
  \langle W_{\tau_{m}}\rangle=\frac{k\tau_{m}}{k_{B}T}\frac{Ak(\tau + \tau_{e})}{\tau_{e}(\gamma_{0}\tau + \gamma\tau_{e} + \gamma_{0}\tau_{e} + k\tau_{e}^{2} + k\tau\tau_{e})}
  \label{eq13a}
\end{equation}
The mean rate of work done ($\langle W_{\tau_m} \rangle/\tau_m$) can further be expressed in terms of $\theta (=\frac{A/\tau_e^2}{k_BT/\gamma_0})$ as,
\begin{equation}
    \label{eq:ve_meanwr_th}
    \frac{\langle W_{\tau_m} \rangle}{\tau_m} = \frac{k^2 \tau_e}{\gamma_0}\frac{\theta (1+\tau/\tau_e)}{[\gamma + \gamma_0(1+\tau/\tau_e) + k\tau_e(1+\tau/\tau_e)]}. 
\end{equation} 
The corresponding expression for the viscous environment can be obtained by taking $\tau \rightarrow 0$ limit as
\begin{equation}
    \label{eq:vis_meanwr_th}
     \left[\frac{\langle W_{\tau_m} \rangle}{\tau_m}\right]_ {\tau \rightarrow 0} = \frac{k^2 \tau_e}{\gamma_0}\frac{\theta}{(\gamma+\gamma_0+k\tau_e)}\equiv \frac{Ak^2}{k_BT}\frac{1}{\tau_e(\gamma+\gamma_0+k\tau_e)}, 
\end{equation}
where $(\gamma + \gamma_0)$ is the effective viscosity of the viscous fluid.
\\

Now the variance of the work is the following:
\begin{align}
\begin{split}
 \text{Var} (W_{\tau_{m}})=\langle(W_{\tau_{m}} - \langle W_{\tau_{m}}\rangle)^{2}\rangle =\left(\frac{k}{k_{B}T}\right)^{2}\Bigg[\tau_{m}\Delta t \text{ACF}_{x_{0}\Dot{x}}(0)&+ 2\tau_{m}\int_{\Delta t}^{\tau_{m}}\text{ACF}_{x_{0}\Dot{x}}(t')dt'\\&- 2\int_{\Delta t}^{\tau_{m}}t'\text{ACF}_{x_{0}\Dot{x}}(t')dt'\Bigg],
 \end{split}
 \label{eq14a}
\end{align}
where, $\Delta t$ is the time step between two observations and $\text{ACF}_{x_{0}\Dot{x}}(t)$ is the auto-correlation function of the product of $x_{0}(t)$ and $\Dot{x}(t)$, i.e., $\text{ACF}_{x_{0}\Dot{x}}(t)=\langle x_{0}(t)\Dot{x}(t) - x_{0}(0)\Dot{x}(0)\rangle$. $\text{ACF}_{x_{0}\Dot{x}}(t)$ can be related to the auto and cross correlation functions of $x_{0}(t)$ and $x(t)$, i.e., $\text{ACF}_{x}$, $\text{ACF}_{x_{0}}$, $\text{XCF}_{x,x_{0}}$ and $\text{XCF}_{x_{0},x}$, which is described in the subsection below.\\

\subsection*{Relating $\text{ACF}_{x_{0}\Dot{x}}$ to $\text{ACF}_{x}$, $\text{ACF}_{x_{0}}$, $\text{XCF}_{x,x_{0}}$ and $\text{XCF}_{x_{0},x}$:}

To find the auto-correlation function of $x_{0}(t)\Dot{x}(t)$, note that the steady-state probability density function of $\boldsymbol{Y}$ (see Eq.~\eqref{eq7a}) is a multivariate Gaussian distribution. If $\boldsymbol{\chi}$ is a multivariate Gaussian distribution with components $\chi_{i}$, $\chi_{j}$, $\chi_{k}$ and $\chi_{l}$ then $$\langle \chi_{i}\chi_{j}\chi_{k}\chi_{l}\rangle=\text{Cov}(\chi_{i},\chi_{j})\text{Cov}(\chi_{k},\chi_{l})+\text{Cov}(\chi_{i},\chi_{k})\text{Cov}(\chi_{j},\chi_{l})+\text{Cov}(\chi_{i},\chi_{l})\text{Cov}(\chi_{j},\chi_{k}).$$
Therefore, we can write 
\begin{align}
\begin{split}
  \langle x_{0n}x_{n}x_{0(n+p)}x_{(n+p)}\rangle=\text{Cov}(x_{0n},x_{n})\text{Cov}(x_{0(n+p)},x_{(n+p)})&+\text{Cov}(x_{0n},x_{0(n+p)})\text{Cov}(x_{n},x_{(n+p)})\\&+\text{Cov}(x_{0n},x_{(n+p)})\text{Cov}(x_{n},x_{0(n+p)}),
\end{split}
\label{eq15a}
\end{align}
where $x_{0n}$ and $x_{n}$ are the n-th points of $x_{0}$ and $x$ time series respectively, and "Cov" represents covariance. Now, note that the auto-correlation function of $x_{0}(t)x(t)$ is represented in the discrete form for lag-time $p\Delta t$ as $\left[\langle x_{0n}x_{n}x_{0(n+p)}x_{(n+p)}\rangle - \langle x_{0n}x_{n}\rangle^{2}\right]$. In the continuous form, from Eq.~\eqref{eq15a},  we have the autocorrelation function of $x_{0}(t)x(t)$,
\begin{equation}
    \text{ACF}_{x_{0}x}(t)=\text{ACF}_{x_{0}}(t)\text{ACF}_{x}(t)+\text{XCF}_{x_{0},x}(t)\text{XCF}_{x,x_{0}}(t),
    \label{eq16a}
\end{equation}
where, "XCF" represents cross-correlation function.
Similarly, we can write in the discrete form,
\begin{align}
    \text{ACF}_{x_{0}\Dot{x}}(p\Delta t)&=\langle x_{0n}\Dot{x}_{n}x_{0(n+p)}\Dot{x}_{(n+p)}\rangle - \langle x_{0n}\Dot{x}_{n}\rangle^{2}\\\nonumber&=\Bigg\langle x_{0n}\left(\frac{x_{n}-x_{(n-1)}}{\Delta t}\right)x_{0(n+p)}\left(\frac{x_{(n+p)}-x_{(n-1+p)}}{\Delta t}\right)\Bigg\rangle - \Bigg\langle x_{0n}\left(\frac{x_{n}-x_{(n-1)}}{\Delta t}\right)\Bigg\rangle^{2}
    \label{eq17a}
\end{align}
when $\Delta t\rightarrow0$. After rearranging carefully and using Eq.~\eqref{eq16a}, we get the expression of the auto-correlation function of $x_{0}(t)\Dot{x}(t)$ as the following:
\begin{equation}
    \text{ACF}_{x_{0}\Dot{x}}(t)=\text{ACF}_{x_{0}}(t)\text{ACF}_{\Dot{x}}(t) - \text{XCF}_{x_{0},\Dot{x}}(t)\text{XCF}_{\Dot{x},x_{0}}(t).
    \label{eq18a}
\end{equation}
Where,
\begin{equation}
    \text{ACF}_{\Dot{x}}(t)=-\frac{d^{2}}{dt^{2}}\{\text{ACF}_{x}(t)\},
    \label{eq19a}
\end{equation}
\begin{equation}
    \text{XCF}_{x_{0},\Dot{x}}(t)=\frac{d\{\text{XCF}_{x_{0},x}(t)\}}{dt},
    \label{eq20a}
\end{equation}
and
\begin{equation}
    \text{XCF}_{\Dot{x},x_{0}}(t)=\frac{d\{\text{XCF}_{x,x_{0}}(t)\}}{dt}.
    \label{eq21a}
\end{equation}
Further, from Eq.~\eqref{eq8a}, we have, $\text{ACF}_{x}(t)=C_{11}(t)$, $\text{XCF}_{x_{0},\Dot{x}}(t)=C_{31}(t)$ and $\text{XCF}_{\Dot{x},x_{0}}(t)=C_{13}(t)$. Therefore the corresponding expressions can be derived from the correlation matrix $\boldsymbol{C}$.

\subsection{Short-time work probability density function}
\label{ap:short_time_wf}
We can write Eq.~\eqref{eq10a} in the discrete form as 
\begin{equation}
    W_{\tau_{m}}=\frac{k}{k_{B}T}\sum_{i=1}^{n(=\tau_{m}/\Delta t)}x_{0i}\Dot{x}_{i}\Delta t = \sum_{i=1}^{n(=\tau_{m}/\Delta t)}W_{i}.
\label{eq:w_taum}
\end{equation}
where $W_{i}$ is the i-th point of the short-time work $W$, which is defined as
\begin{equation}
    W_{i}=\frac{k}{k_{B}T}x_{0i}\Dot{x}_{i}\Delta t.
\label{eq:w_short}
\end{equation}
Now, note that $x_{0}$ and $\Dot{x}\Delta t$ are two Gaussian distributed cross-correlated quantities which are given by $\mathcal{N}(0,\sigma_{33}^{2})$ and  $\mathcal{N}(0,\Delta t^{2}\text{ACF}_{\Dot{x}}(0))$ respectively with the normalized cross-correlation $\rho=\frac{\text{XCF}_{x_{0},\Dot{x}}(0)}{\sigma_{33}\sqrt{\text{ACF}_{\Dot{x}}(0)}}$. Therefore, the probability density function of $z=x_{0}\Dot{x}\Delta t$, can be calculated from the joint probability density function $f$ of these two quantities from the relation \cite{cui2016exact}
\begin{equation}
    \text{P}(z)=\int_{-\infty}^{\infty}\frac{1}{\left|y\right|}f(y,z/y)dy.
    \label{eq:P_z_s}
\end{equation}
Here,
\begin{align}
\label{eq:ff}
    f(x_{0},\Dot{x}\Delta t)&=\frac{1}{2\pi\sigma_{33}\sqrt{\Delta t^{2}\text{ACF}_{\Dot{x}}(0)}\sqrt{1-\rho^{2}}}\times\\\nonumber
    &\exp\left[-\frac{1}{2(1-\rho^{2})}\left(\frac{x_{0}^{2}}{\sigma_{33}^{2}} - \frac{2\rho x_{0}\Dot{x}\Delta t}{\sigma_{33}\sqrt{\Delta t^{2}\text{ACF}_{\Dot{x}}(0)}} + \frac{(\Dot{x}\Delta t)^{2}}{\Delta t^{2}\text{ACF}_{\Dot{x}}(0)}   \right)  \right].
\end{align}
Using Eq.~\eqref{eq:ff} for Eq.~\eqref{eq:P_z_s}, we get,
\begin{align}
    P(z)=\frac{\exp\left(\frac{\rho z}{\sigma_{33}\sqrt{\Delta t^{2}\text{ACF}_{\Dot{x}}(0)}(1-\rho^{2})}\right)}{2\pi\sigma_{33}\sqrt{\Delta t^{2}\text{ACF}_{\Dot{x}}(0)}\sqrt{1-\rho^{2}}}\int_{-\infty}^{\infty}\frac{1}{|y|}\exp\left[-\frac{1}{2(1-\rho^{2})}\left( \frac{y^{2}}{\sigma_{33}^{2}} + \frac{z^{2}}{\Delta t^{2}\text{ACF}_{\Dot{x}}(0)y^{2}}\right)\right]dy.
    \label{eq:pz}
\end{align}
The integral can be solved using the formula given in \cite{gradshteyn2014table}, and we calculate,
\begin{align}
    P(z)=\mathcal{A}\exp(\beta z)\text{K}_{0}(\alpha|z|),
    \label{eq:pz_final}
\end{align}
where, $\mathcal{A}=\frac{1}{\pi\sigma_{33}\sqrt{\Delta t^{2}\text{ACF}_{\Dot{x}}(0)}\sqrt{1-\rho^{2}}}$, $\beta=\frac{\rho}{\sigma_{33}\sqrt{\Delta t^{2}\text{ACF}_{\Dot{x}}(0)}(1-\rho^{2})}$, $\alpha=\frac{1}{\sigma_{33}\sqrt{\Delta t^{2}\text{ACF}_{\Dot{x}}(0)}(1-\rho^{2})}$ and $\text{K}_{0}$ is the zeroth order modified Bessel function. The moment-generating function corresponding to $P(z)$ is 
\begin{align}
\label{eq:mom_s}
    \text{M}_{1}(s)&=\mathcal{A}\int_{-\infty}^{\infty}e^{sz}e^{\beta z}\text{K}_{0}(\alpha|z|)dz\\\nonumber&=\mathcal{A}\pi\frac{1}{\sqrt{\alpha^{2} - (s+\beta)^{2}}}
\end{align}

From the normalization condition, $\text{M}_{1}(0)=1$, we find a relation between $\mathcal{A}$ and $\alpha$, $\beta$.
Now to take into account the factor $\frac{k}{k_{B}T}$ in the Eq.~\eqref{eq:w_short}, defining $\alpha_{ws}=\frac{\alpha}{k/k_{B}T}$, $\beta_{ws}=\frac{\beta}{k/k_{B}T}$ and $\mathcal{A}_{ws}=\frac{\sqrt{\alpha_{ws}^{2} - \beta_{ws}^{2}}}{\pi}$, we finally obtain the exact expression of the short-time work probability density function
\begin{align}    
P(W)=\mathcal{A}_{ws}\exp(\beta_{ws} W)\text{K}_{0}(\alpha_{ws}|W|).
    \label{eq:pws_final}
\end{align}
The related moment generating function will be exactly the same as given in the Eq.~\eqref{eq:mom_s} after replacing $\mathcal{A}$ and $\alpha$, $\beta$ by $\mathcal{A}_{ws}$ and $\alpha_{ws}$, $\beta_{ws}$.

\subsection{Cumulative work probability density function} 
\label{ap:cumulative_wf}
Assuming the time-series of $W$ uncorrelated, we can say $W_{\tau_{m}}$ is just the summation over the independent points from that time-series of $W$ as defined in the Eq.~\eqref{eq:w_taum}. Therefore, the corresponding moment generating function will be the following:
\begin{equation}
    \text{M}_n(s)=\left(\frac{\mathcal{A}_{ws}\pi}{\sqrt{\alpha_{ws}^{2} - (s+\beta_{ws})^{2}}}\right)^{n}.
    \label{eq:mom_l}
\end{equation}
Interestingly, the characteristic function of this moment-generating function can be written as
\begin{equation}
    \text{CF}(s)=\left(\frac{\mathcal{A}_{ws}\pi}{\sqrt{\alpha_{ws}^{2} - \beta_{ws}^{2}}}\right)^{n}\left(1-\frac{is}{\alpha_{ws} - \beta_{ws}}\right)^{-\frac{n}{2}}\left(1+\frac{is}{\alpha_{ws} + \beta_{ws}}\right)^{-\frac{n}{2}}.
    \label{cf_l}
\end{equation}
The inverse fourier transform of $\text{CF}(s)$ is the probability density function of the sum of $n$ number of $W$ points. Now, note that $\left(1-\frac{is}{\alpha_{ws} - \beta_{ws}}\right)^{-\frac{n}{2}}\left(1+\frac{is}{\alpha_{ws} + \beta_{ws}}\right)^{-\frac{n}{2}}$ is related to the CF of the probability density function of $\text{y}=\text{y}_{1} - \text{y}_{2}$ where $\text{y}_{1}$ and $\text{y}_{2}$ are two independent numbers drawn from two Gamma distributions $\Gamma(\frac{n}{2},\alpha - \beta)$ and $\Gamma(\frac{n}{2},\alpha + \beta)$ respectively \cite{mathai1993noncentral,gradshteyn2014table}. Therefore, the desired probability density function (for the uncorrelated W) is the convolution of two Gamma distributions and is given by
\begin{equation}
    \text{P}(W_{\tau_{m}})=\frac{\mathcal{A}_{ws}^{n}\pi^{n}}{\sqrt{\pi}\Gamma(\frac{n}{2})}\frac{1}{(2\alpha_{ws})^{\frac{n-1}{2}}}e^{\beta_{ws} Z}|W_{\tau_{m}}|^{\frac{n-1}{2}}\text{K}_{\frac{n-1}{2}}(\alpha_{ws}|W_{\tau_{m}}|),
    \label{eq:unco_wpdf_l}
\end{equation}
where $\text{K}_{\frac{n-1}{2}}$ is the modified Bessel function of order $(n-1)/2$.
Further, it can be shown that $\alpha_{ws}$ and $\beta_{ws}$ are also related to the mean and variance of $W_{\tau_{m}}$ from the moment generating function Eq.~\eqref{eq:mom_l} such that 
\begin{equation}
\begin{split}
   & \langle W_{\tau_m}\rangle = \frac{d}{ds}[\text{M}_n(s)]_{s=0}= \frac{n\beta_{ws}}{\alpha_{ws}^2-\beta_{ws}^2}, \\
   & \text{Var}(\text{W}_{\tau_{m}}) = \frac{d^2}{ds^2}[\text{M}_n(s)]_{s=0}= \frac{n(\alpha_{ws}^2+\beta_{ws}^2)}{\alpha_{ws}^2-\beta_{ws}^2}.
    \end{split}
\end{equation}

Using these two relations, $\beta_{ws}$ and $\alpha_{ws}$ can be calculated as,
\begin{equation}
\begin{split}
    &\beta_{ws}=\frac {n\langle\text{W}_{\tau_{m}}\rangle}{n\text{Var} (\text{W}_{\tau_{m}})-2\langle\text{W}_{\tau_{m}}\rangle^2},\\
    &\alpha_{ws}=\left[\beta_{ws}^{2} + \frac{n\beta_{ws}}{\langle \text{W}_{\tau_{m}}\rangle}\right]^{\frac{1}{2}}.
    \label{eq:alpha_beta_rel}
    \end{split}
\end{equation}
Note that Eq.~\eqref{eq:unco_wpdf_l} would converge to the short-time work probability distribution function for $n\rightarrow 1$. 

Unfortunately, the time-series of $W$ is not uncorrelated, and hence, Eq.~\eqref{eq:unco_wpdf_l} does not describe the experimentally and numerically measured data. However, we have found an approximate expression of the cumulative work probability distribution function. Assuming that the correlation will not change the form of $P(W_{\tau_{m}})$ given in the Eq.~\eqref{eq:unco_wpdf_l}, we can use the mean and variance of $W_{\tau_{m}}$ calculated directly from the Langevin equation, i.e., Eqns~\eqref{eq13a} and \eqref{eq14a} respectively, to calculate the $\alpha$ and $\beta$ parameters. Notably, Eqns~\eqref{eq13a} and \eqref{eq14a} do not neglect the correlation in $W$. We call these new parameters $\alpha_{w}$ and $\beta_{w}$, and finally reach the approximate expression of the cumulative work probability distribution function which is the following:
\begin{equation}
    \text{P}(W_{\tau_{m}})=\frac{\mathcal{A}_{w}^{n}\pi^{n}}{\sqrt{\pi}\Gamma(\frac{n}{2})}\frac{1}{(2\alpha_{w})^{\frac{n-1}{2}}}e^{\beta_{w} Z}|W_{\tau_{m}}|^{\frac{n-1}{2}}\text{K}_{\frac{n-1}{2}}(\alpha_{w}|W_{\tau_{m}}|).
    \label{eq:co_wpdf_l}
\end{equation}
where, $\mathcal{A}_{w}=\frac{\sqrt{\alpha_{w}^{2}-\beta_{w}^{2}}}{\pi}$. We have found that this approximate expression of $P(W_{\tau_{m}})$ matches the experimental and numerical data excellently which is described in the main text.

\section{Work fluctuation theorem}
\label{ap:wft}
\begin{figure}[ht]
    \centering
    \includegraphics[width = 0.90\textwidth]{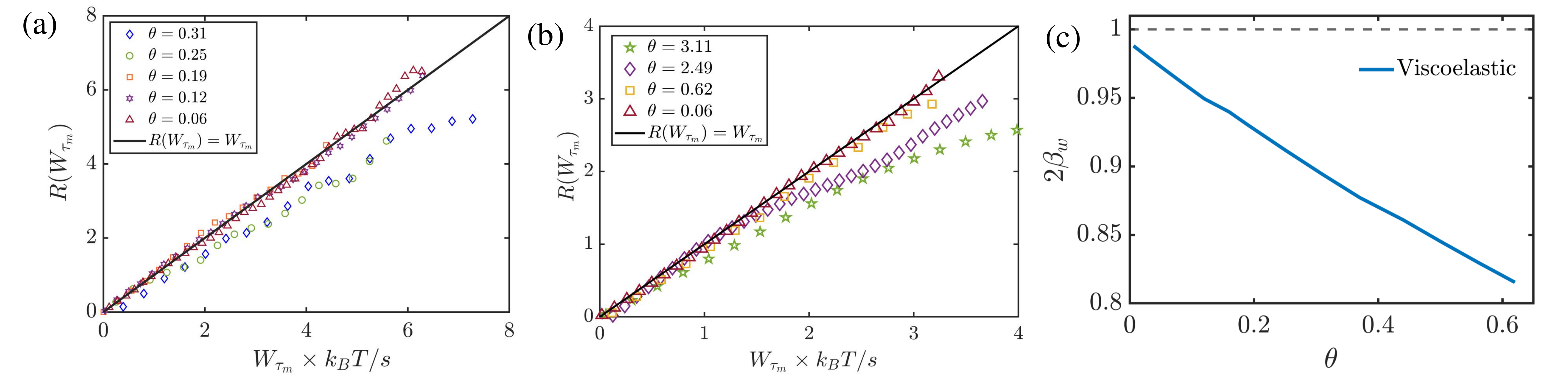}
    \caption{Ratio of probabilities corresponding to the positive and negative work fluctuations is plotted for different strengths of external noise in the viscoelastic (a) and viscous (b) bath, respectively.(c) Plot of $2\beta_{w}$ as a function of $\theta$.} 
    \label{fig:ap_wft}
\end{figure}

Here we show the ratio of probabilities corresponding to the positive and negative fluctuations of the stochastic work done on a microscopic particle in the viscoelastic (Fig.~\ref{fig:ap_wft}(a)) and viscous bath (Fig.~\ref{fig:ap_wft}(b)),-- estimated from the numerical data. As discussed in the main text, $R(W_{\tau_m})$ for the viscous bath follows the unit slope line for higher strength of the external noise than the same in the viscoelastic bath. The slope, however, as expected, converges to one as $\theta$ tends to zero (Fig.~\ref{fig:ap_wft}(c)).

\section{ Entropy production rate}
\label{ap:EPR}
The entropy production rate of this system can be calculated using the Fokker-plank based method as described in Ref.~\cite{li2019quantifying,manikandan2021quantitative,das2022inferring,dey2022experimental}. Starting from an arbitrary initial condition for $x, X$ and $x_0$, the system will reach to a nonequilibrium steady state in the long time with a characteristics probability distribution ($P_{ss}(\mathbf{Y})$) and current ($\mathbf{J}_{ss}(\mathbf{Y})$) that can be described in terms of the steady state covariance matrix $\boldsymbol{\sigma}$ as shown in  Eq.\eqref{eq:prob_current1}.

Now the total entropy production rate (in units of $k_B s^{-1}$) of the system can be obtained as,
 \begin{equation}
    \label{eq:epr_form}
    \Pi = \int d \mathbf{Y} \frac{\mathbf{D}_{f}^{-1} \mathbf{J}_{ss}^2 (\mathbf{Y})}{P_{ss}(\mathbf{Y})}  =  \frac{Ak^2(\tau + \tau_{e})}{T\tau_{e}(\gamma_{0}\tau + \gamma\tau_{e} + \gamma_{0}\tau_{e} + k\tau_{e}^{2} + k\tau\tau_{e})}. 
\end{equation}

Note that this expression is exactly similar to the mean rate of work done ($\langle W_{\tau_m}\rangle/\tau_m$) as shown in Eq.\eqref{eq13a}.

\section{Method to obtain the rheological properties of the fluid}
We deploy a phase-sensitive active microrheological technique to measure the fluid characteristics of the viscoelastic medium \cite{paul2019active}. In this technique, an optically trapped microscopic particle embedded in the fluid is perturbed by an external square modulation containing sinusoids of several frequencies, and the position fluctuations of the particle are recorded. From the recorded trajectory, the relative phase of the particle's response to the modulation is measured as a function of the modulation frequency. The relaxation timescale and the relative viscosity of the fluid can be directly estimated by fitting the phase response curve with the theoretical expression calculated from the Langevin description of the trapped particle in the viscoelastic fluid, following Jeffrey's model. The overdamped dynamics of the  particle trapped in an optical potential with stiffness $k$ can be written in the frequency domain (taking Fourier transform of the overdamped version of Eq. \eqref{eq1}) as:
\begin{equation}
    i\omega \Gamma(\omega)x(\omega) + kx(\omega) = kx_0(\omega)
    \label{eq:dyn_freq}
\end{equation}
where, 
\begin{align}
\Gamma(\omega) = \int_{0}^{\infty} \Gamma(t) \exp(i\omega t)\ dt = \Bigg[\gamma_0 + \frac{\gamma}{-i\omega \tau +1 }\Bigg]\\
= 6\pi a_0\mu_s\Bigg[1 + \frac{\mu_r}{-i\omega \tau +1 }\Bigg].
\label{eq:drag_freq}
\end{align}
The relative viscosity $\mu_r$ is the ratio of corresponding viscosities of the polymer ($\mu_p$) and the solvent($\mu_s$). $a_0$ denotes the radius of the trapped particle. Now Eq.\ref{eq:drag_freq} can be substituted in Eq.\ref{eq:dyn_freq} to get the phase response of the particle as:
\begin{equation}
    \delta(\omega) = \tan^{-1}\Bigg[\frac{\frac{1+\mu_r}{\tau^2}\omega + \omega^3}{\frac{k}{\tau^2\gamma_0}+(\frac{k}{\gamma_0}-\frac{\mu_r}{\tau})\omega^2}\Bigg].
    \label{eq:phase_response}
\end{equation}
As described in Ref.\cite{paul2019active}, the trapped microparticle embedded in the viscoelastic fluid is perturbed with a square wave modulation of frequency $3 \ Hz$.
The phase response of the  particle to the odd harmonics of the square wave modulation is  obtained experimentally and fitted with  Eq.\ref{eq:phase_response} to estimate the relaxation timescale ($\tau$), and relative viscosity ($\mu_r$) of the fluid  as shown in Fig.\ref{fig:diagram}(b) of the main text.  

\label{ap:rheology}
\begin{figure}[h]
    \centering
    \includegraphics[width = 0.75\textwidth]{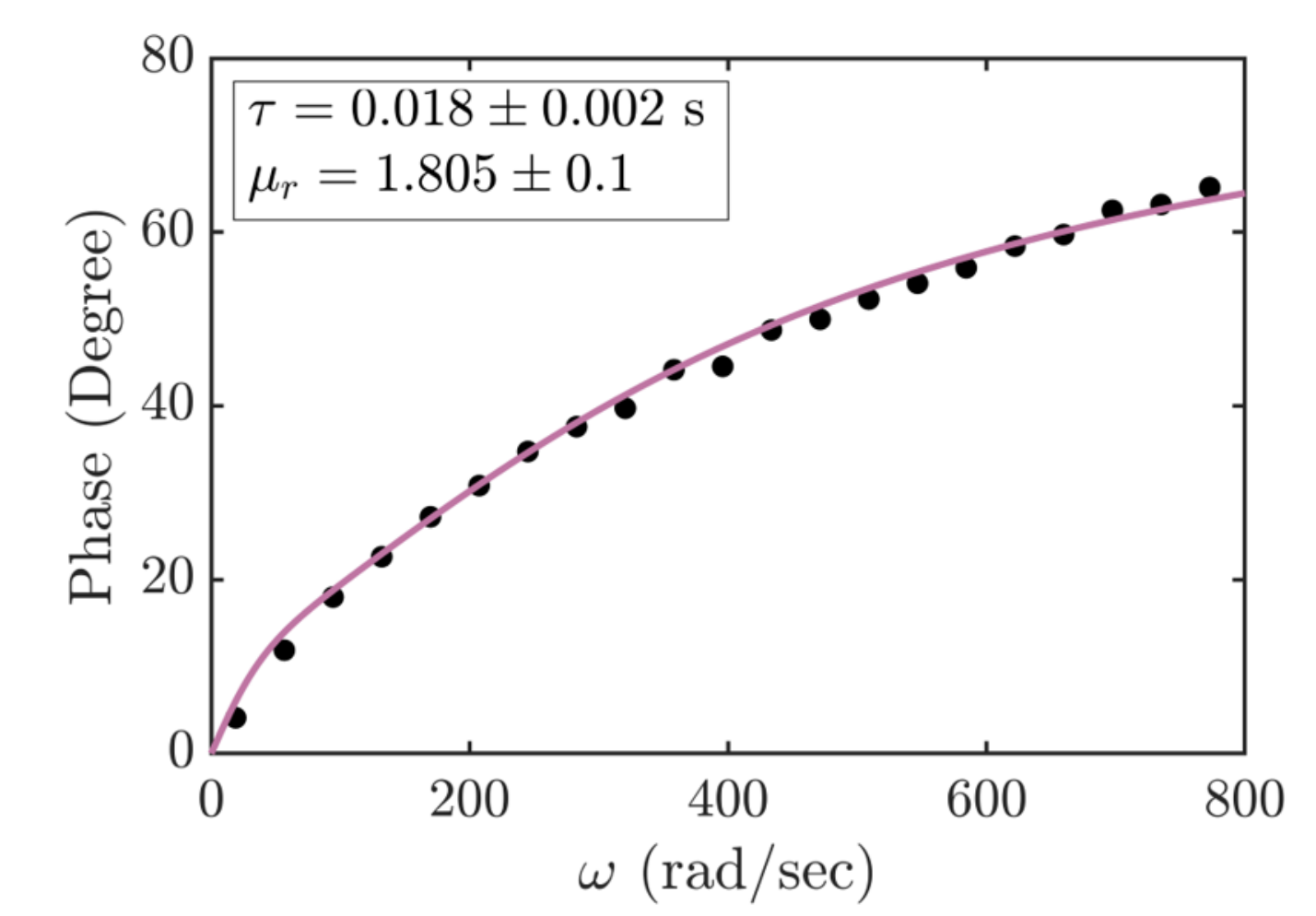}
    \caption{Phase response of the trapped particle for different frequencies and the corresponding fit for the phase-sensitive active micro-rheology  \cite{paul2019active}. }
    \label{fig:rheology_measure}
\end{figure}

\section{Experimental setup}
\label{ap:exp}
\begin{figure}
    \centering
    \includegraphics[width = 0.85\textwidth]{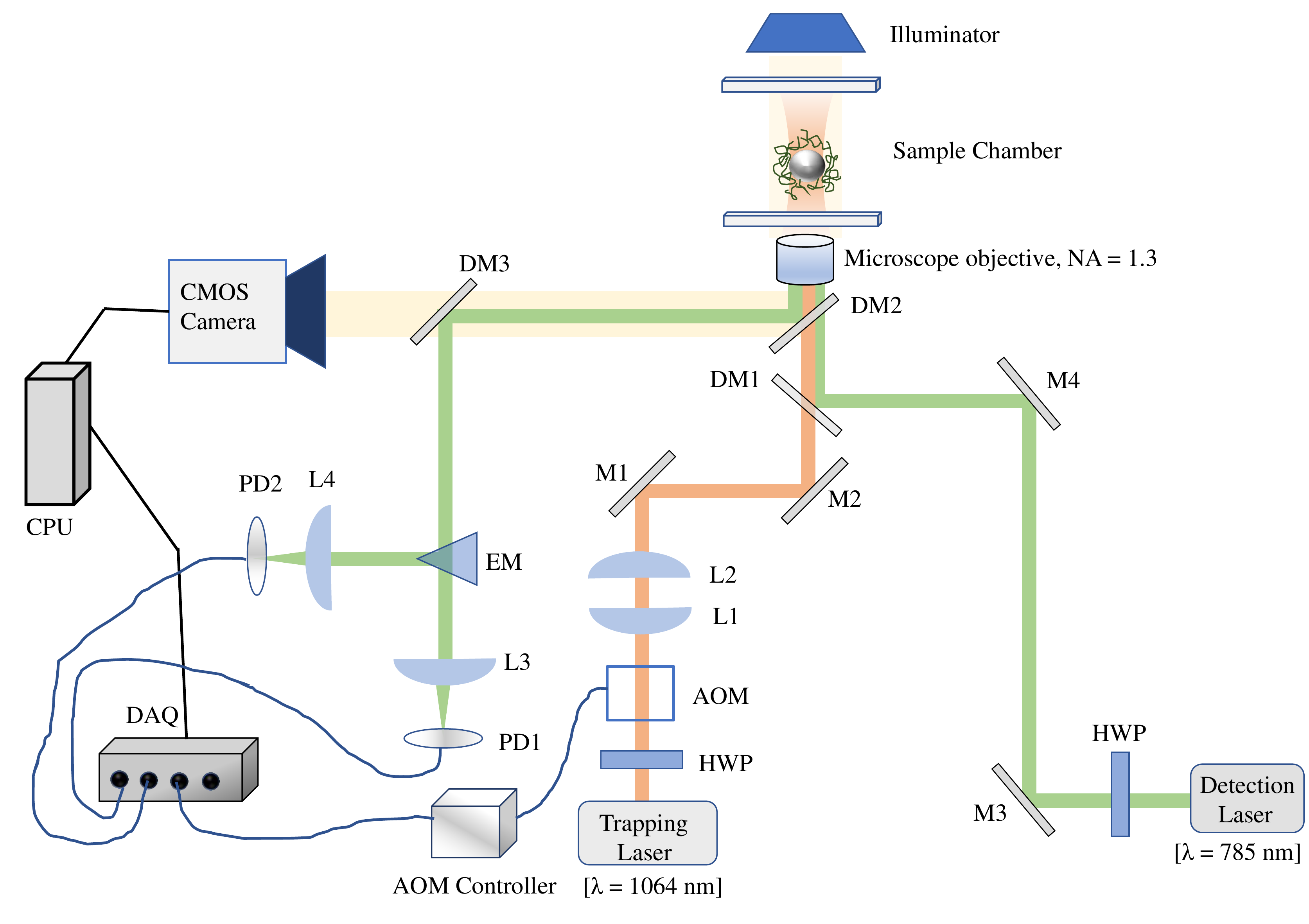}
    \caption{Schematic of our setup. The annotations are as follows: M: Mirror, DM: Dichroic mirror, EM: Edge mirror, L: Plano-convex lens of different focal lengths, HWP: Half-wave plate, PD: Photodiode, AOM: Acousto optic modulator, DAQ: Data acquisition card (National Instruments).  }
    \label{fig:setup_schematic}
\end{figure}

Our Optical tweezers setup (Fig.~\ref{fig:setup_schematic}) is developed around a standard inverted microscope (Olympus IX71) with a high numerical aperture objective lens (100X, NA = 1.3). A semiconductor laser (Lasever, Max power 500 mW) of wavelength 1064 nm is tightly focused through the objective lens on the sample to trap a micro-particle dispersed in it. The trapping beam is passed through an acousto-optic modulator (AOM) placed at a plane conjugate to the objective lens. We modulate the beam by using the first-order diffracted beam off the AOM with a computer-generated signal fed to the AOM controller via a data acquisition card (which acts as a signal generator here). We employ a second low-power laser beam (stationary and copropagating with the trapping beam) of wavelength 785 nm to track the position of the particle, which we determine from the back-scattered light incident on a balanced detection system \cite{bera2017fast}, constructed using two high gain-bandwidth photo-detectors (Thorlabs PDA100A2) placed orthogonal to each other. The displacement data of the particle is recorded into a computer from the balanced detection signal via a data acquisition card (NI DAQ) at a spatio-temporal resolution of 1 nm-10kHz. 

\section*{References}
%\bibliographystyle{unsrt}
%\bibliography{references}

\end{document}